\documentclass[aps,prb,reprint,superscriptaddress]{revtex4-2}

\usepackage{graphicx}
\usepackage{bbm}
\usepackage{braket}
\usepackage{amsmath}
\usepackage{amsfonts}
\usepackage{graphicx}
\usepackage{amssymb}
\usepackage{color}
\usepackage{hyperref}
\usepackage{listings}
\usepackage{array}
\usepackage{physics}

\usepackage[usenames,dvipsnames]{xcolor}

\begin{document}
\title{Formation, prevalence, and stability of bouncing-ball quantum scars}
\author{Simo Selinummi}
\email{simo.selinummi@tuni.fi}
\affiliation{Computational Physics Laboratory, Tampere University, P.O. Box 600, FI-33014 Tampere, Finland}
\author{Joonas Keski-Rahkonen}
\affiliation{Computational Physics Laboratory, Tampere University, P.O. Box 600, FI-33014 Tampere, Finland}
\affiliation{Department of Physics, Harvard University, Cambridge, Massachusetts 02138, USA}
\affiliation{Department of Chemistry and Chemical Biology, Harvard University, Cambridge, Massachusetts 02138, USA}
\author{Fartash Chalangari}
\affiliation{Computational Physics Laboratory, Tampere University, P.O. Box 600, FI-33014 Tampere, Finland}
\author{Esa R\"as\"anen}
\affiliation{Computational Physics Laboratory, Tampere University, P.O. Box 600, FI-33014 Tampere, Finland}

\begin{abstract}

Quantum scars correspond to enhanced probability densities along unstable classical periodic orbits. In recent years, research on quantum scars has extended to various systems including the many-body regime. In this work we focus on the formation, prevalence, and stability of linear “bouncing-ball” (BB) scars in two-dimensional (2D) quantum wells. These scars have relevance as effective and controllable channels in quantum transport. We utilize imaginary time propagation to solve the 2D Schrödinger equation within an arbitrary external confining potential, specifically the quantum well model with external perturbations. We show how BB scars begin to emerge with a single perturbative peak, such as a repulsive bump or attractive dip that simulates the effect of a charged nanotip in the system. We then identify the optimal size of the perturbative peak to maximize the prevalence of these scars. Finally, we investigate the stability of BB scars against external noise and find that some of them are remarkably robust. This suggests promising opportunities for further applications of BB scars in quantum transport.

\end{abstract}

\maketitle

\section{Introduction}

A quantum eigenstate of a classically chaotic system is formally defined to be scarred by a periodic orbit if its density on the classical invariant manifolds near and along that periodic orbit is systematically enhanced beyond the classically expected density.~\cite{Heller_book} This phenomenon, first discovered, named, and explained by Heller in 1984, presents a counterintuitive manifestation of both quantum-classical correspondence and the quantum suppression of classical chaos.~\cite{Heller_phys.rev.lett_53_1515_1984} Without scarring, one might expect the eigenstate probability densities of a classically chaotic system to uniformly cover position space up to quantum fluctuations, reflecting ergodicity of classical chaos. Gutzwiller's trace formula~\cite{Gutzwiller_j.math.phys_12_343_1971, Gutzwiller_book} supports this expectation, suggesting no single periodic orbit dominates the quantum density of states. However, contrary to this misleading assumption, scarred states flourish in the gaps left by the quantum ergodicity theorems of Shnirelman~\cite{Shnirelman_Uspekhi.Mat.Nauk_29_181_1974}, de Verdiere~\cite{Colindeverdiere_comm.math.phys_102_497_1985}, and Zelditch~\cite{zelditch_duke.math.j_55_919_1987}. Along with the theoretical advancements~\cite{kaplan_ann.phys_264_171_1998, Kaplan_nonlinearity_12_R1_1999, bogomolny_physica.d_31_169_1988, berry_proc_r_soc_lond_a_423_219_1989, kus_phys.rev.a_43_4244_1991, dariano_phys.rev.a_45_3646_1992, tomsovic_phys.rev.lett_70_1405_1993, revuelta_phys.rev.e_102_042210_2020,agam_j.phys.a.math_26_2113_1993,bohigas_phys.rep_223_43_1993, wisniacki_phys.rev.lett_97_094101_2006}, experimental evidence of scars has nowadays accumulated within a vast selection of experiments~\cite{Fromhol_phys.rev.lett_75_1142_1995, Wilkinson_nature_380_608_1996, narimanov_phys.rev.lett_80_49_1998, Honig_phys.rev.a_39_5642_1989, bogomolny_phys.rev.lett_97_254102_2006, kim_phys.rev.b_65_165317_2002, stockman_phys.rev.lett.64.2215_1990, dorr_phys.rev.lett_80_1030_1998, Sridhaar_phys.rev.lett_67_785_1991, Stein_phys.rev.lett_68_2867_1992, nockel_nature_385_45_1997, Lee_phys.rev.lett_88_033903_2002, Harayama_phys.rev.e_67_015207_2003, chinnery_phys.rev.e_53_272_1996}. 

In addition to the conventional scarring according to Heller's original work~\cite{Heller_phys.rev.lett_53_1515_1984}, the concept of quantum scars has recently expanded into three distinct areas: relativistic, many-body, and variational. While scars in relativistic quantum systems share the same origins as the conventional scarring~\cite{Huang_phys.rev.lett_103_054101_2009}, they differ in features such as their recurrence with energy variation and the potential to exhibit chirality~\cite{Xu_phys.rev.lett_110_064102_2013, xuan_phys.rev.E_86_016702_2012, song_phys_rev.research_1_033008_2019}. Many-body scars, on the other hand, are special states in many-body Hilbert space that evade thermalization at finite energy densities ~\cite{bernien_nature_551_579_2017,scherg_nat.commun_12_1_2021, zhao_phys.rev.lett_124_160604_2020}, recently linked to conventional scarring. ~\cite{Hummel_phys.rev.Lett_130_250402_2023, Evrard_phys.rev.lett_132_020401_2024}. These states cause persistent oscillations of local observables without relying on (near)integrability or the protection provided by a global symmetry~\cite{turner_nat.phys_14_745_2018, Ho_phys.rev.lett_122_040603_2019, serbyn_nat.phys_17_675_2021}. The third category is variational (or perturbation-induced) scarring~\cite{Luukko_sci.rep_6_37656_2016, keski-rahkonen_phys.rev.b_97_094204_2017, keski-rahkonen_j.phys.conden.matter_31_105301_2019, keski-rahkonen_phys.rev.lett_123_214101_2019, Luukko_phys.rev.lett_119_203001_2017, antiscarring}. This type of scarring resembles conventional scarring but emerges as a result of the combined effects of perturbations and near-degeneracies in the unperturbed system, as determined by the variational principle, which is why it is referred to as perturbation-induced.

Within the classification of quantum scars, a fascinating subspecies is the probability density condensation along an orbit linked to the bouncing motion of a classical particle. In the context of a stadium-shaped billiard, these highly non-ergodic "bouncing-ball" (BB) eigenstates are restricted to vertical bouncing motions between the straight walls and can be shown~\cite{OConnor_phys.rev.lett_61_2288_1988} to persist even at infinite energy. This finding indicates that while they survive at high energies, they represent an increasingly smaller fraction of the total number of states, aligning with the quantum ergodicity theorem. Similar BB scars also arise in the context of variational scarring within a disordered quantum dot~\cite{keski-rahkonen_phys.rev.b_97_094204_2017, keski-rahkonen_phys.rev.lett_123_214101_2019}, attributed to unperturbed states with negligible angular momentum. In these systems, BB scars are defined as the quantum equivalent of radial, linear back-and-forth motion.

In this paper, we explore variational BB scars in a two-dimensional (2D) quantum well (QW). First, we confirm earlier findings that strongly scarred eigenstates, which display features of BB motion, can be generated by a single, localized perturbation, such as a potential bump or dip created by a nanotip. A key aspect of our study is the extension of the analysis to include both positive and negative perturbations in the potential. We also clarify how these BB scars emerge from degenerate sets of states in the weak-perturbation limit. Next, we show that the strength and frequency of these scars can be optimized by tailoring the scar-generating perturbation, enabling control over the scar's orientation. Additionally, we find that these scars are robust against noise from other perturbations, such as impurities.

\section{Model system}\label{SecThGLLB}

We consider a 2D QW described by the following generic single-electron Hamiltonian in atomic units:
\begin{equation}\label{Hamiltonian}
H = \frac{1}{2}\nabla^2 + V_{\textrm{ext}} + V_{\textrm{tip}} + V_{\textrm{imp}},
\end{equation}
composed of the external confinement $V_{\textrm{ext}}$, perturbing nanotip $V_{\textrm{tip}}$ and impurity noise $V_{\textrm{imp}}$. 
This kind of Hamiltonian is directly relevant for modeling semiconductor QWs influenced by impurities (see, e.g., Refs.~\cite{Halonen_Phys.Rev.B_53_6971_1996, Rasanen_Phys.Rev.B_70_115308_2004, Guclu_Phys.Rev.B_68_035304_2003}). It serves as an excellent platform for studying quantum chaos by comparing it to classical billiards with realistic soft walls. This comparison involves a statistical analysis of energy levels, scarring, and ergodicity.~\cite{keski-rahkonen_j.phys.conden.matter_31_105301_2019} Additionally, these types of open QWs are suitable for wave function imaging based on energy shifts in single-particle resonances induced by an atomic force microscopy tip~\cite{mendoza_phys.rev.b_68_205302_2003, zozoulenko_phys.rev.b_5810597_1998}, enabling indirect observation of conductance fluctuations resulting from scarred states. Moreover, the scarred eigenstates of an electron in a QW can be measured via quantum tomography (see, e.g., Ref.~\cite{jullien2_Nature_514_603_2014}) or directly mapped with scanning tunneling microscopy, as presented in Ref.~\cite{ge2024direct}.

We have here chosen the external potential $V_{\textrm{ext}} = \frac{1}{2}r^5$, which serves as a prototypical framework for investigating variational scarring; it represents a natural "sweet spot" for studying this phenomenon. In the absence of impurities, the confinement of the QW also establishes the energy scales and the integrability of the system. Notably, the geometry of a periodic orbit (PO) is independent of its energy in our QW, which belongs to a specific class of homogeneous potentials. As a result, different POs can be easily enumerated using just two integers:~\cite{keski-rahkonen_phys.rev.b_97_094204_2017} after $a$ oscillations around the radial turning points, the particle has traveled around the origin $b$ times before returning to its original configuration. The most notable PO is a five-pointed star: the orbit closes on itself after two rounds around the origin ($a$ = 2) during five radial oscillations ($b$ = 5)~\cite{Luukko_sci.rep_6_37656_2016}. Additionally, there are two special POs: circular orbits (which involve no radial motion) and BBs (which have no angular momentum), the latter being the central focus of this study.

On the quantum side, the eigenstates $\vert r, m\rangle$ of the unperturbed, circularly symmetric system are labeled by two quantum numbers $(r, m)$, corresponding to radial and angular motion, respectively. In the absence of a magnetic field, the states $\vert r, \pm m\rangle$ are exactly degenerate. Moreover, there are also near-degeneracies, or quasi-degeneracies, intimately connected to classical POs. According to the Einstein-Brillouin-Keller quantization rules, with fixed Maslov indices leading to a Bohr-Sommerfeld-like consideration, if a state defined by quantum numbers $(r, m)$ is nearby in action to a classical PO with a ratio $a/b$ in the radial and angular oscillation frequencies, the relative states $\vert r + ka, m - kb\rangle$, where $k \in \mathbf{N}$, will consequently be nearby in energy. These groups of nearly-degenerate states are informally known as a resonant set, which, along with exact degeneracies, constitutes the first component of the scar recipe.

\begin{figure}
\centering
\includegraphics[width=1.0 \linewidth]{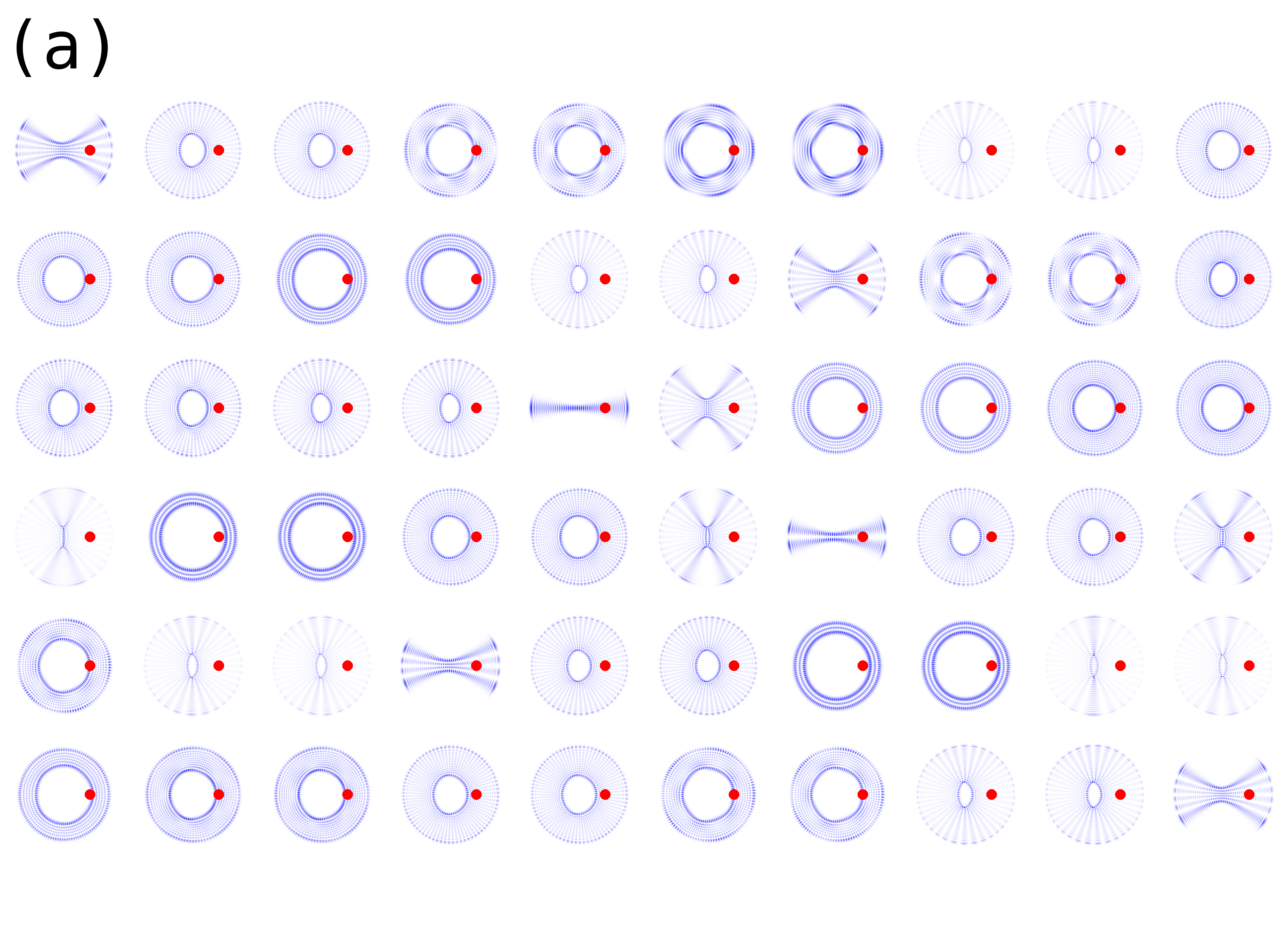}
\label{fig:states}
\hfill
\centering
\includegraphics[width=0.49 \linewidth]{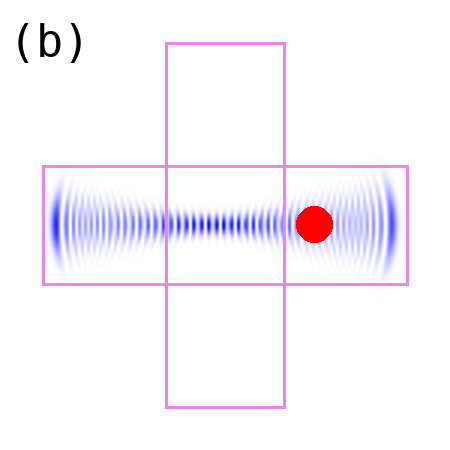}
\label{fig:pinned_bb}
\hfill
\centering
\includegraphics[width=0.49 \linewidth]{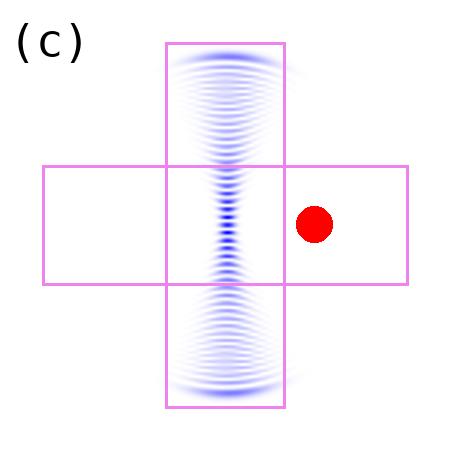}
\label{fig:avoided_bb}
\caption{(a) Examples of eigenstates in a 2D quantum well perturbed by a potential bump with $A_T = 50$ and $\sigma_T = 1.5$. (b), (c) Rectangular areas where the normalized probability densities are calculated in order to detect BB scars. In (b) the perturbation parameters are the same as in (a) but in (c), the perturbation is a potential dip with $A_T = -50$ and $\sigma_T = 1.5$.} 
\label{fig:bb_box_examples}
\end{figure}

The second essential component for variational scarring involves a perturbation applied to the system. The perturbation induced by a nanotip takes the form of a potential bump or dip centered at location $\mathbf{r}_0$, expressed as
\begin{equation}
    V_{\textrm{tip}} = A_{T} \exp\left({\frac{\vert \mathbf{r} - \mathbf{r}_0 \vert^2}{2\sigma_T^2}}\right),
\end{equation}
where $A_T$ and $\sigma_T$ denote the amplitude and width of the nanotip perturbation. This type of Gaussian profile is a well-validated approximation for local perturbations caused by a conducting nanotip. When sufficiently strong, the nanotip generates a distinct set of scarred eigenstates in the perturbed system out of a specific resonant set. Due to the connection between the states in the resonant set and classical motion, some linear combinations of the resonant set will lead to an interference pattern tracing out a path of a classical PO. Furthermore, because of the spatially localized nature of the perturbation, these scarred states are favored according to the variational principle: scars showcased in Fig.~\ref{fig:bb_box_examples} can effectively maximize or minimize the perturbation by orienting so that they either coincide with or avoid the nanotip, respectively. These two scenarios for a BB scar are displayed in Fig.~\ref{fig:bb_box_examples}.

Besides the nanotip, real QW devices are often affected by impurities and imperfections. This class of perturbation can be modeled by adding randomly located bumps to the otherwise smooth confining potential, yielding the total noise
\begin{equation}
    V_{\textrm{imp}} = A_{N} \sum_i \exp\left({\frac{\vert \mathbf{r} - \mathbf{r}_i \vert^2}{2\sigma_N^2}}\right),
\end{equation}
where the impurities are uniformly distributed throughout the confining potential, with an average density of one impurity per unit square. Similarly to the perturbation caused by the nanotip, the individual bumps of the noise are assumed to be Gaussian-like with amplitude $A_N$ and width $\sigma_N$, but restricted to individually weaker perturbations than the nanotip. For simplicity, we further assume the impurity perturbations to share the same amplitude and width. This kind of disorder mode has been studied with density-functional theory~\cite{hirose_phys.rev.B_65_193305_2002, hirose_phys.Rev.B_63_075301_2001} and the diffusive quantum Monte Carlo approach~\cite{Guclu_Phys.Rev.B_68_035304_2003}. Moreover, the role of such impurities within a QW can be quantitatively identified through the measured differential magnetoconductance displaying the quantum eigenstates~\cite{Rasanen_Phys.Rev.B_70_115308_2004}.

\section{Results}\label{SecApplGLLB}

To compute the first 3025 eigenstates of the Hamiltonian (\ref{Hamiltonian}), we utilize $\texttt{itp2d}$ software \cite{LuukkoP.J.J.2013Itpc} that utilizes the imaginary time propagation (ITP) algorithm as it is particularly effective for 2D problems. A Gaussian potential induces a perturbation in the confining potential: a bump with a positive amplitude or a potential dip with a negative amplitude. As illustrated in Fig.~\ref{fig:bb_box_examples}, the perturbation is positioned along the x-axis of the 2D plane in the location (2,0). Due to the variational principle, the BB scars are either horizontally oriented to pin themselves to the perturbation or vertically oriented to avoid the perturbation. These differently oriented BB scars are then detected by integrating the amount of normalized probability densities inside rectangular horizontal and vertical areas which are shown in Fig. \ref{fig:bb_box_examples}. By establishing thresholds for the probability densities, we can distinguish BB scars from other types of scars and analyze their emergence.

\subsection{Formation of bouncing-ball scars}

\begin{figure*}
\centering
\includegraphics[width=\textwidth]{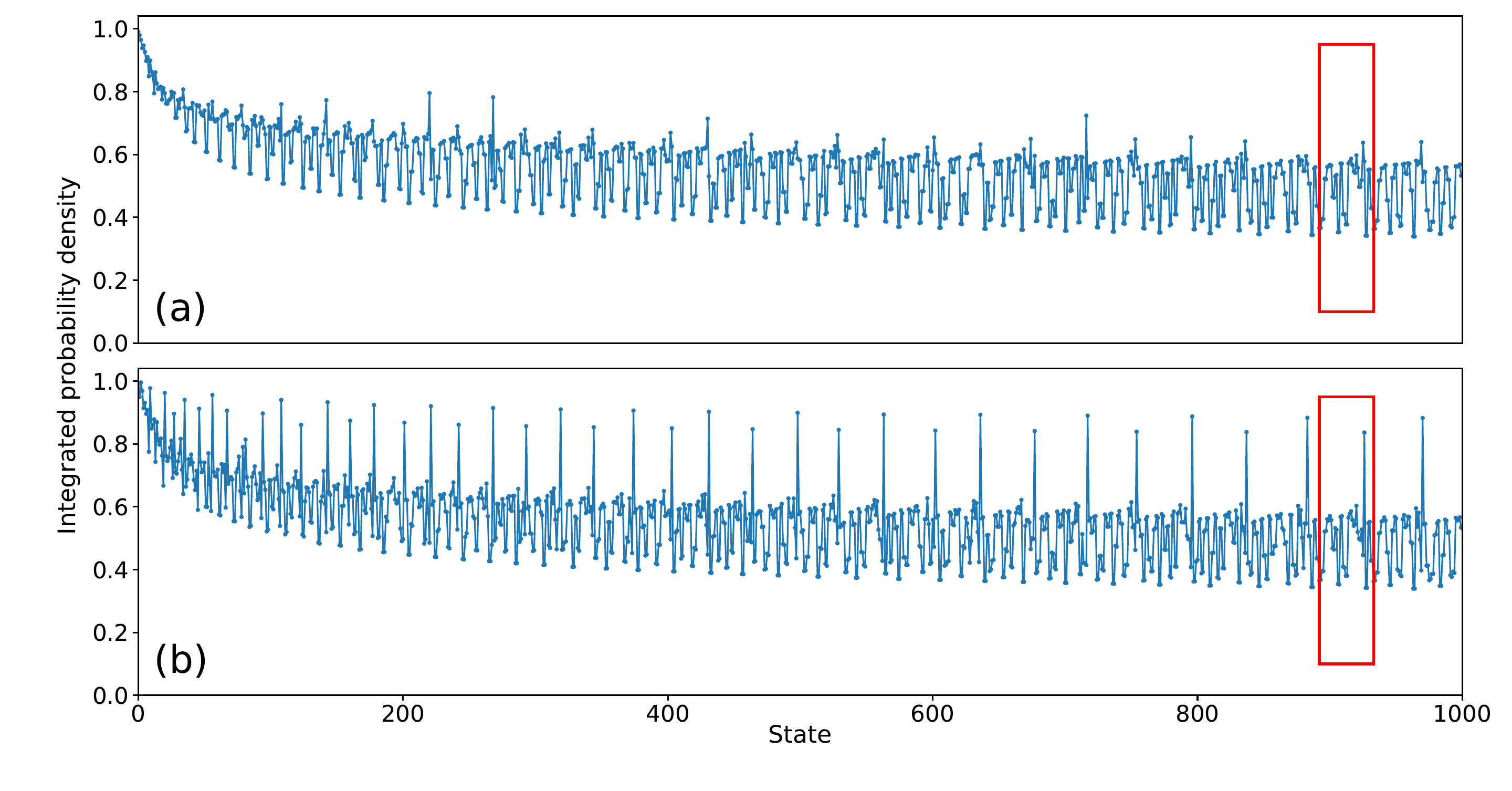}
\caption{Comparison of the integrated probability densities of states between a clean system (a) and a system perturbed by a repulsive bump with $A_T = 4$ and $\sigma_T = 0.235$ (b).}
\label{fig:density_comparison}
\end{figure*}

In order to further understand the development of BB scarring in a rotationally symmetric system under perturbation, we first analyze the weak-perturbation limit. Specifically, we focus on the behavior of the integrated probability density in the horizontal BB area depicted in Fig. \ref{fig:bb_box_examples}(b) across the eigenstates. Figure \ref{fig:density_comparison} compares the integrated probability densities across the spectra of a clean system (a) and a perturbed system (b) for the first 1000 eigenstates. The perturbation is a repulsive bump with parameters $A_T = 4$ and $\sigma_T = 0.235$. We find clear spikes in (b) when a perturbation is included, indicating increased concentration within the detection box. These spikes appear in a repeating pattern, resulting from transitions between degenerate resonant sets of eigenstates. One such resonant set is highlighted with red boxes in Fig. \ref{fig:density_comparison}. Similar behavior can also be observed with an attractive dip for the case of a vertical detection box.

\begin{figure*}
\centering
\includegraphics[width=\textwidth]{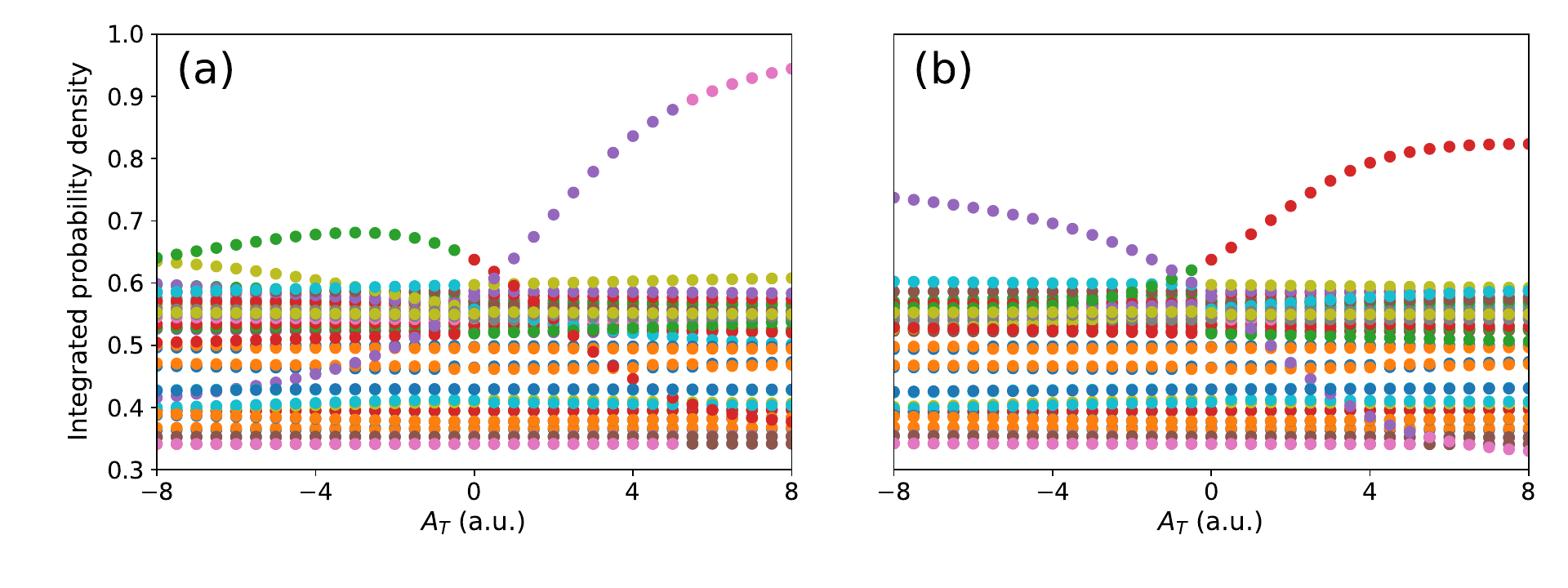}
\caption{Integrated probability densities of the degenerate resonant set states (see Fig. \ref{fig:density_comparison}) in (a) horizontal and (b) vertical detection boxes that overlap with the perturbation; see Fig. \ref{fig:bb_box_examples}(b). As the amplitude of the perturbation is varied some eigenstates linked to bouncing-ball scars within the resonant set are more significantly influenced by the presence of a perturbation.}
\label{fig:affected_states}
\end{figure*}

Figure~\ref{fig:affected_states} offers an in-depth view of the integrated probability densities of the resonant set highlighted in Fig. \ref{fig:density_comparison}. In this analysis, the amplitude of the perturbation varies around zero, transitioning from an attractive dip (negative $A_T$) to a repulsive bump (positive $A_T$), while the perturbation width is fixed at $\sigma=0.235$. In Fig. \ref{fig:affected_states}(a), the densities are integrated over the horizontal detection box, and in (b), over the vertical box. The colors in the graphs represent different state numbers within the resonant set. As illustrated, most states are only slightly affected by the perturbation's presence. However, certain states—associated with BB scars—are highly responsive to it, leading to significant changes in the regions where their probability densities are concentrated. The probability densities increasingly occupy the detection boxes, indicating the emergence of BB scars.

\subsection{Prevalence of different bouncing-ball scars}

The prevalence of BB scars can be assessed by calculating their occurrences among all the eigenstates using two differently oriented detection boxes, as illustrated in Fig. \ref{fig:bb_box_examples}, and applying a threshold for the integrated probability density. To concentrate on relatively strong BB scars, we establish a threshold of 0.90, meaning that over $90\%$ of the probability density must be localized within the box. A total of 3025 eigenstates are solved for the model system. However, we exclude the first 300 eigenstates because they are primarily confined to the center of the system (within the detection box) and therefore contribute to the scar count despite their BB characteristics.  For illustration, the first 300 states are included in Fig.~\ref{fig:density_comparison} above.

\begin{figure}
\centering
\includegraphics[width=0.45\textwidth]{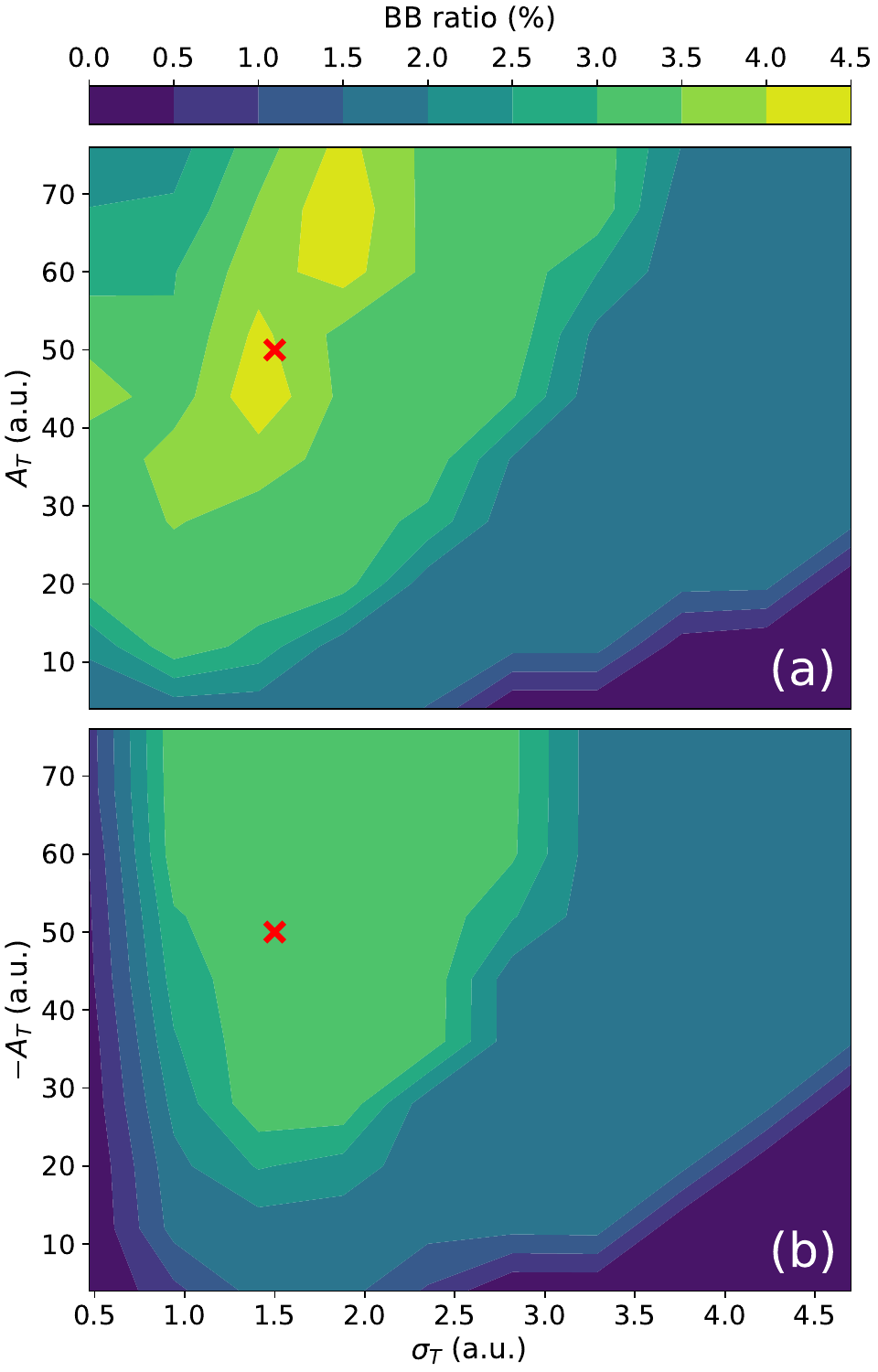}
\caption{Proportion of BB scars with different perturbation parameter combinations. A repulsive bump is used as the perturbation in (a) and an attractive dip in (b).}
\label{fig:colormaps}
\end{figure}

Figure~\ref{fig:colormaps} elucidates the effect of the perturbation parameters onto the prevalence of BB scars in the model system. The prevalence is here defined as the ratio of detected BB scars to the number of solved eigenstates. As shown, distinct regions in the figure indicate that by adjusting the amplitude $A_T$ or width $\sigma_T$ of the perturbation, the number of BB scars can be controlled and maximized. This effect is evident for both perturbation types, as depicted in Figs. \ref{fig:colormaps}(a) and (b), which represent the cases of a repulsive bump and an attractive dip, respectively. However, a comparison of the two figures indicates that the efficiency of BB scar induction varies between the two types of perturbations. Counterintuitively, the repulsive bump proves to be more effective, requiring smaller parameters to achieve a higher prevalence of BB scars compared to the attractive dip. Moreover, the maximum number of BB scars is attained with the bump rather than the dip. Nevertheless, this outcome is reasonable, as a bump creates a repulsive region that pushes states away, potentially distorting or scattering them more significantly than a dip, which may trap some states locally but does not disrupt the overall spectrum as much. Therefore, in relative terms, a positive perturbation (bump) can be interpreted as a stronger perturbation than a negative one.

\begin{figure}
\centering
\includegraphics[width=0.5\textwidth]{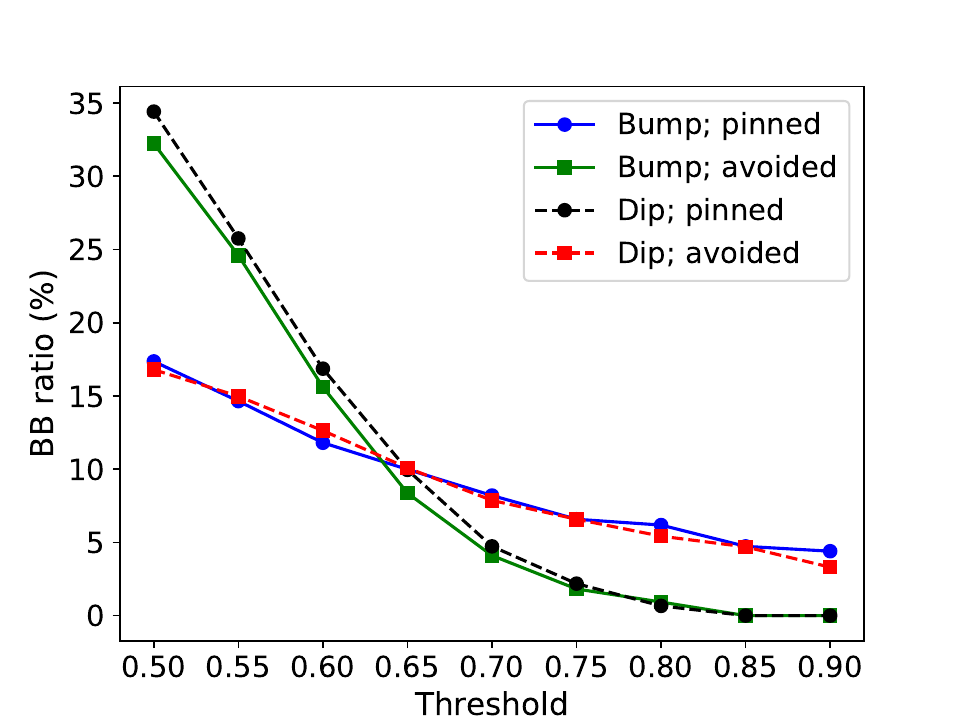}
\caption{Bouncing-ball scar ratios for different perturbation types and scar orientations. Amplitude $A_T = 50$ is used for the bump and $A_T = -50$ for the dip, with the width being $\sigma_T = 1.5$ for both.}
\label{fig:bb_ratio_plot}
\end{figure}

As described in Sec. \ref{SecThGLLB}, BB scars can orient themselves in two different ways with our single perturbation configuration. For experimental applications, it is of importance to understand how these orientations manifest with both perturbation types. Figure~\ref{fig:bb_ratio_plot} displays the prevalence of these two orientations for each perturbation type. A perturbation parameter combination of $A_T = 50$ and $\sigma_T = 1.5$ is used for the bump, while $A_T = -50$ and $\sigma_T = 1.5$ are used for the dip. These parameters yield the highest proportion of BB scars, marked by red crosses in Fig. \ref{fig:colormaps}. The BB ratio is shown as a function of detection box threshold to observe changes in orientation when more spread-out BB scars are included. Figure~\ref{fig:bb_ratio_plot} reveals clear orientation preferences for both perturbation types: strong BB scars tend to pin themselves to the bump, while they avoid the dip. On the other hand, these preferred orientations reverse when more spread-out BB scars are considered.

\subsection{Robustness against noise}
\begin{figure}
\vspace*{.3cm}
\renewcommand{\arraystretch}{0}
\setlength{\tabcolsep}{0pt}
\setlength{\arrayrulewidth}{0.5mm}
\begin{tabular}{|c|c|}
\hline
\includegraphics[width=0.5 \linewidth]{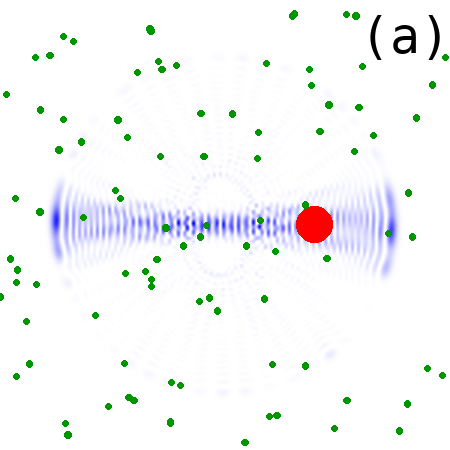} &
\label{fig:A5noise}
\includegraphics[width=0.5 \linewidth]{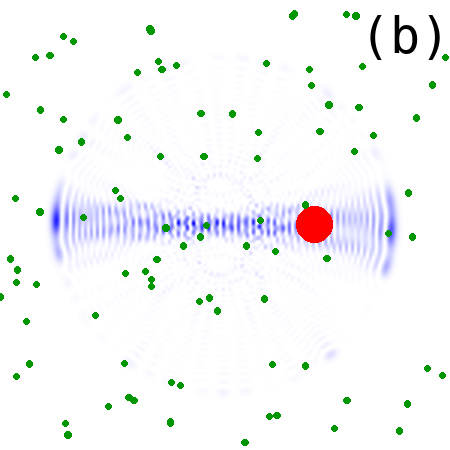}
\label{fig:A6noise} \\
\hline
\includegraphics[width=0.5 \linewidth]{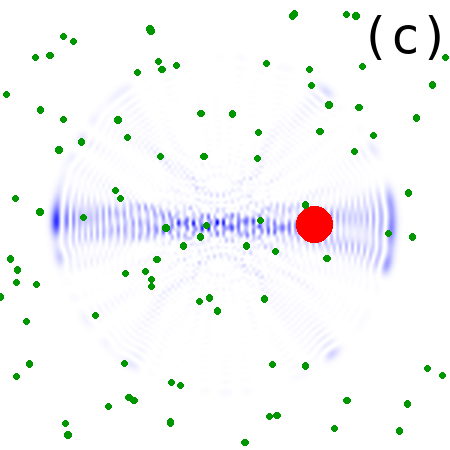} &
\label{fig:A7noise}
\includegraphics[width=0.5 \linewidth]{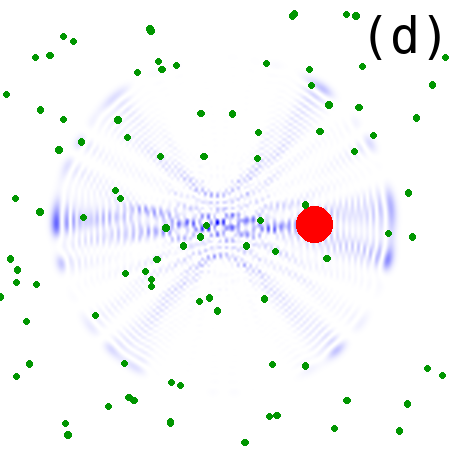}
\label{fig:A8noise} \\
\hline
\end{tabular}
\caption{Demonstration of the degradation of a bouncing-ball scar in state $n = 2167$ as the amplitude of the external noise (green dots) increases from (a) five to (b) six, (c) seven, and (d) eight. The repulsive bump (red circle) has $A_T=50$ and $\sigma_T=1.5$.}
\label{fig:bb_vanishment}
\end{figure}

Physical systems often contain unwanted impurities, making it essential for BB scars to be robust in such environments. Figure~\ref{fig:bb_vanishment} demonstrates the gradual degradation of a single BB scar in state $n=2167$ as noise increases. The primary perturbation is a repulsive bump with $A_T=50$ and $\sigma_T=1.5$, while the noise amplitude starts at $A_N=5$ in panel (a) and increases in increments of one in each subsequent panel. As shown, the BB scar gradually spreads out, and the probability density is no longer concentrated within either detection box. However, not all BB scars disperse in the same manner; some remain within the detection box for a longer duration as noise levels rise.

\begin{figure}[ht!]
\centering
\includegraphics[width=0.5\textwidth]{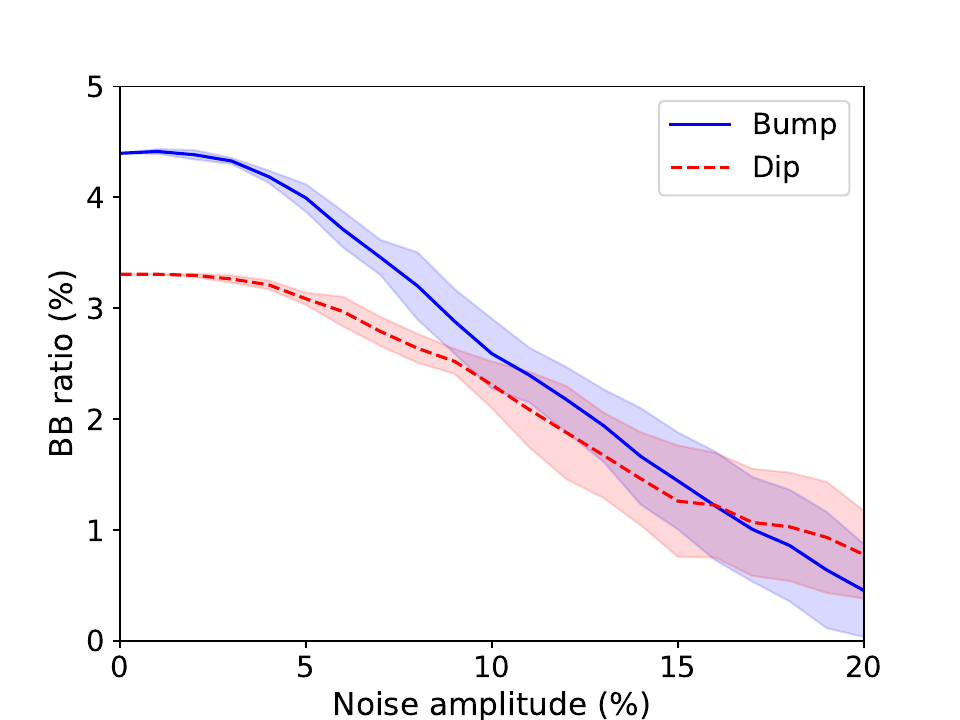}
\caption{Relative proportion of bouncing-ball scars out of all eigenstates as a function of noise amplitude, presented as percentages of the main perturbation. The lines correspond to averages of ten different noise location configurations, and the shaded areas around the lines illustrate standard deviation.}
\label{fig:bb_decrease_avg}
\end{figure}

The impact of noise on the prevalence of BB scars can be statistically examined by calculating the BB ratios as noise levels increase. We again use perturbation parameters $A_T = 50$ for the primary bump and $A_T = -50$ for the primary dip, with the width being $\sigma_T = 1.5$, as these values yield a high proportion of BB scars. Next, we introduce repulsive bumps as noise, which are randomly distributed within the potential well. The seed for these random locations is a controllable variable, enabling comparisons of the same location configurations under different noise strengths. The amplitude-to-width ratio of the noise bumps is the same as in  the primary bumps and dips.

Figure~\ref{fig:bb_decrease_avg} depicts the effect of noise on the prevalence of BB scars, with BB ratios calculated for ten different noise location configurations. The graphs display averages across these configurations, and the shaded areas indicate standard deviations. The results reveal that there is a minimal decline in BB ratios until the noise amplitude reaches $4\,\%$ of the main perturbation amplitude, after which the ratios gradually decrease as the noise amplitude increases. The critical amplitude threshold of $4\,\%$  likely results from the selected width of the detection box: as noise increases, the strong BB scars begin to spread out and fill the detection box, eventually leaking out at around $4\,\%$. Nevertheless, even after exceeding this threshold, the decline remains gradual, highlighting the resilience of BB scars. A similar robustness was observed with attractive noise bumps serving as noise (not shown), exhibiting behavior closely resembling that in Fig. \ref{fig:bb_decrease_avg}.

An interesting phenomenon occurs when the noise amplitude approaches approximately $20\,\%$. At this point, the high-amplitude noise bumps significantly disrupt the system, leading to a diminished effect of the main perturbation. This combined influence creates new BB scars, which may realign themselves in response to the noise bumps. In other words, multiple orientations can now optimize the perturbation based on the variational principle, resulting in a loss of control over the orientation of the BB scars.

\section{Conclusions}\label{SecConc}

In summary, we have demonstrated that a two-dimensional quantum well, perturbed by a single Gaussian bump or dip in the confining potential—simulating the effect of a quantum nanotip—leads to the formation of linearly shaped bouncing-ball scars. We explicitly highlight the pathway that some of the states in the degenerate set take to transform into BB scars, with their orientations determined by the variational principle. Our findings reveal that the prevalence of BB scars can be controlled by tuning the perturbation parameters, enabling the maximization of scar formation with carefully chosen characteristics. Specifically, a repulsive Gaussian bump generates strong, spatially localized BB scars, while an attractive dip mainly causes the scars to avoid the perturbation. Additionally, we have demonstrated that BB scars are robust against external noise, maintaining their structure even under perturbative disruptions. The ability to design and stabilize these scars suggests potential applications for BB scars as controllable channels in quantum transport, opening new possibilities for exploiting quantum states even in impurity-plagued nanostructures.
\\

\begin{acknowledgements}
This work was funded by the Research Council of Finland, ManyBody2D Project No. 349956. The authors acknowledge CSC-IT Center for Science, Finland, for generous computational resources. Furthermore, J.K.-R. thanks the Oskar Huttunen Foundation for financial support.
\end{acknowledgements}

\bibliography{Bibliography}

\begin{thebibliography}{61}%
\makeatletter
\providecommand \@ifxundefined [1]{%
 \@ifx{#1\undefined}
}%
\providecommand \@ifnum [1]{%
 \ifnum #1\expandafter \@firstoftwo
 \else \expandafter \@secondoftwo
 \fi
}%
\providecommand \@ifx [1]{%
 \ifx #1\expandafter \@firstoftwo
 \else \expandafter \@secondoftwo
 \fi
}%
\providecommand \natexlab [1]{#1}%
\providecommand \enquote  [1]{``#1''}%
\providecommand \bibnamefont  [1]{#1}%
\providecommand \bibfnamefont [1]{#1}%
\providecommand \citenamefont [1]{#1}%
\providecommand \href@noop [0]{\@secondoftwo}%
\providecommand \href [0]{\begingroup \@sanitize@url \@href}%
\providecommand \@href[1]{\@@startlink{#1}\@@href}%
\providecommand \@@href[1]{\endgroup#1\@@endlink}%
\providecommand \@sanitize@url [0]{\catcode `\\12\catcode `\$12\catcode
  `\&12\catcode `\#12\catcode `\^12\catcode `\_12\catcode `\%12\relax}%
\providecommand \@@startlink[1]{}%
\providecommand \@@endlink[0]{}%
\providecommand \url  [0]{\begingroup\@sanitize@url \@url }%
\providecommand \@url [1]{\endgroup\@href {#1}{\urlprefix }}%
\providecommand \urlprefix  [0]{URL }%
\providecommand \Eprint [0]{\href }%
\providecommand \doibase [0]{https://doi.org/}%
\providecommand \selectlanguage [0]{\@gobble}%
\providecommand \bibinfo  [0]{\@secondoftwo}%
\providecommand \bibfield  [0]{\@secondoftwo}%
\providecommand \translation [1]{[#1]}%
\providecommand \BibitemOpen [0]{}%
\providecommand \bibitemStop [0]{}%
\providecommand \bibitemNoStop [0]{.\EOS\space}%
\providecommand \EOS [0]{\spacefactor3000\relax}%
\providecommand \BibitemShut  [1]{\csname bibitem#1\endcsname}%
\let\auto@bib@innerbib\@empty
\bibitem [{\citenamefont {Heller}(2018)}]{Heller_book}%
  \BibitemOpen
  \bibfield  {author} {\bibinfo {author} {\bibfnamefont {E.~J.}\ \bibnamefont
  {Heller}},\ }\href@noop {} {\emph {\bibinfo {title} {The Semiclassical Way to
  Dynamics and Spectroscopy}}}\ (\bibinfo  {publisher} {Princeton University
  Press},\ \bibinfo {year} {2018})\BibitemShut {NoStop}%
\bibitem [{\citenamefont {Heller}(1984)}]{Heller_phys.rev.lett_53_1515_1984}%
  \BibitemOpen
  \bibfield  {author} {\bibinfo {author} {\bibfnamefont {E.~J.}\ \bibnamefont
  {Heller}},\ }\bibfield  {title} {\bibinfo {title} {Bound-state eigenfunctions
  of classically chaotic hamiltonian systems: Scars of periodic orbits},\
  }\href {https://doi.org/10.1103/PhysRevLett.53.1515} {\bibfield  {journal}
  {\bibinfo  {journal} {Phys. Rev. Lett.}\ }\textbf {\bibinfo {volume} {53}},\
  \bibinfo {pages} {1515} (\bibinfo {year} {1984})}\BibitemShut {NoStop}%
\bibitem [{\citenamefont
  {Gutzwiller}(1971)}]{Gutzwiller_j.math.phys_12_343_1971}%
  \BibitemOpen
  \bibfield  {author} {\bibinfo {author} {\bibfnamefont {M.~C.}\ \bibnamefont
  {Gutzwiller}},\ }\bibfield  {title} {\bibinfo {title} {Periodic orbits and
  classical quantization conditions},\ }\href
  {https://doi.org/10.1063/1.1665596} {\bibfield  {journal} {\bibinfo
  {journal} {J. Math. Phys}\ }\textbf {\bibinfo {volume} {12}},\ \bibinfo
  {pages} {343} (\bibinfo {year} {1971})}\BibitemShut {NoStop}%
\bibitem [{\citenamefont {Gutzwiller}(1991)}]{Gutzwiller_book}%
  \BibitemOpen
  \bibfield  {author} {\bibinfo {author} {\bibfnamefont {M.}~\bibnamefont
  {Gutzwiller}},\ }\href@noop {} {\emph {\bibinfo {title} {Chaos in Classical
  and Quantum Mechanics}}},\ Interdisciplinary Applied Mathematics\ (\bibinfo
  {publisher} {Springer New York},\ \bibinfo {year} {1991})\BibitemShut
  {NoStop}%
\bibitem [{\citenamefont
  {Shnirel'man}(1974)}]{Shnirelman_Uspekhi.Mat.Nauk_29_181_1974}%
  \BibitemOpen
  \bibfield  {author} {\bibinfo {author} {\bibfnamefont {A.~I.}\ \bibnamefont
  {Shnirel'man}},\ }\bibfield  {title} {\bibinfo {title} {Ergodic properties of
  eigenfunctions},\ }\href {http://www.ams.org/mathscinet-getitem?mr=402834}
  {\bibfield  {journal} {\bibinfo  {journal} {Uspekhi Mat. Nauk}\ }\textbf
  {\bibinfo {volume} {29}},\ \bibinfo {pages} {181} (\bibinfo {year}
  {1974})}\BibitemShut {NoStop}%
\bibitem [{\citenamefont {Colin~de
  Verdi{\'e}re}(1985)}]{Colindeverdiere_comm.math.phys_102_497_1985}%
  \BibitemOpen
  \bibfield  {author} {\bibinfo {author} {\bibfnamefont {Y.}~\bibnamefont
  {Colin~de Verdi{\'e}re}},\ }\bibfield  {title} {\bibinfo {title}
  {Ergodicit{\'e} et fonctions propres du laplacien},\ }\href
  {https://projecteuclid.org:443/euclid.cmp/1104114465} {\bibfield  {journal}
  {\bibinfo  {journal} {Comm. Math. Phys.}\ }\textbf {\bibinfo {volume}
  {102}},\ \bibinfo {pages} {497} (\bibinfo {year} {1985})}\BibitemShut
  {NoStop}%
\bibitem [{\citenamefont {Zelditch}(1987)}]{zelditch_duke.math.j_55_919_1987}%
  \BibitemOpen
  \bibfield  {author} {\bibinfo {author} {\bibfnamefont {S.}~\bibnamefont
  {Zelditch}},\ }\bibfield  {title} {\bibinfo {title} {Uniform distribution of
  eigenfunctions on compact hyperbolic surfaces},\ }\href
  {https://doi.org/10.1215/S0012-7094-87-05546-3} {\bibfield  {journal}
  {\bibinfo  {journal} {Duke Math. J.}\ }\textbf {\bibinfo {volume} {55}},\
  \bibinfo {pages} {919} (\bibinfo {year} {1987})}\BibitemShut {NoStop}%
\bibitem [{\citenamefont {Kaplan}\ and\ \citenamefont
  {Heller}(1998)}]{kaplan_ann.phys_264_171_1998}%
  \BibitemOpen
  \bibfield  {author} {\bibinfo {author} {\bibfnamefont {L.}~\bibnamefont
  {Kaplan}}\ and\ \bibinfo {author} {\bibfnamefont {E.~J.}\ \bibnamefont
  {Heller}},\ }\bibfield  {title} {\bibinfo {title} {Linear and nonlinear
  theory of eigenfunction scars},\ }\href
  {https://doi.org/https://doi.org/10.1006/aphy.1997.5773} {\bibfield
  {journal} {\bibinfo  {journal} {Ann. Phys. (N. Y.)}\ }\textbf {\bibinfo
  {volume} {264}},\ \bibinfo {pages} {171 } (\bibinfo {year}
  {1998})}\BibitemShut {NoStop}%
\bibitem [{\citenamefont {Kaplan}(1999)}]{Kaplan_nonlinearity_12_R1_1999}%
  \BibitemOpen
  \bibfield  {author} {\bibinfo {author} {\bibfnamefont {L.}~\bibnamefont
  {Kaplan}},\ }\bibfield  {title} {\bibinfo {title} {Scars in quantum chaotic
  wavefunctions},\ }\href {https://doi.org/10.1088/0951-7715/12/2/009}
  {\bibfield  {journal} {\bibinfo  {journal} {Nonlinearity}\ }\textbf {\bibinfo
  {volume} {12}},\ \bibinfo {pages} {R1} (\bibinfo {year} {1999})}\BibitemShut
  {NoStop}%
\bibitem [{\citenamefont {Bogomolny"}(1988)}]{bogomolny_physica.d_31_169_1988}%
  \BibitemOpen
  \bibfield  {author} {\bibinfo {author} {\bibfnamefont {E.~B.}\ \bibnamefont
  {Bogomolny"}},\ }\bibfield  {title} {\bibinfo {title} {Smoothed wave
  functions of chaotic quantum systems},\ }\href
  {https://doi.org/https://doi.org/10.1016/0167-2789(88)90075-9} {\bibfield
  {journal} {\bibinfo  {journal} {Physica D: Nonlinear Phenomena}\ }\textbf
  {\bibinfo {volume} {31}},\ \bibinfo {pages} {169 } (\bibinfo {year}
  {1988})}\BibitemShut {NoStop}%
\bibitem [{\citenamefont {Berry}(1989)}]{berry_proc_r_soc_lond_a_423_219_1989}%
  \BibitemOpen
  \bibfield  {author} {\bibinfo {author} {\bibfnamefont {M.~V.}\ \bibnamefont
  {Berry}},\ }\bibfield  {title} {\bibinfo {title} {Quantum scars of classical
  closed orbits in phase space},\ }\href
  {https://doi.org/10.1098/rspa.1989.0052} {\bibfield  {journal} {\bibinfo
  {journal} {Proc. R. Soc. Lond. A}\ }\textbf {\bibinfo {volume} {423}},\
  \bibinfo {pages} {219} (\bibinfo {year} {1989})}\BibitemShut {NoStop}%
\bibitem [{\citenamefont {Ku\ifmmode~\acute{s}\else \'{s}\fi{}}\ \emph
  {et~al.}(1991)\citenamefont {Ku\ifmmode~\acute{s}\else \'{s}\fi{}},
  \citenamefont {Zakrzewski},\ and\ \citenamefont {\ifmmode~\dot{Z}\else
  \.{Z}\fi{}yczkowski}}]{kus_phys.rev.a_43_4244_1991}%
  \BibitemOpen
  \bibfield  {author} {\bibinfo {author} {\bibfnamefont {M.}~\bibnamefont
  {Ku\ifmmode~\acute{s}\else \'{s}\fi{}}}, \bibinfo {author} {\bibfnamefont
  {J.}~\bibnamefont {Zakrzewski}},\ and\ \bibinfo {author} {\bibfnamefont
  {K.}~\bibnamefont {\ifmmode~\dot{Z}\else \.{Z}\fi{}yczkowski}},\ }\bibfield
  {title} {\bibinfo {title} {Quantum scars on a sphere},\ }\href
  {https://doi.org/10.1103/PhysRevA.43.4244} {\bibfield  {journal} {\bibinfo
  {journal} {Phys. Rev. A}\ }\textbf {\bibinfo {volume} {43}},\ \bibinfo
  {pages} {4244} (\bibinfo {year} {1991})}\BibitemShut {NoStop}%
\bibitem [{\citenamefont {D'Ariano}\ \emph {et~al.}(1992)\citenamefont
  {D'Ariano}, \citenamefont {Evangelista},\ and\ \citenamefont
  {Saraceno}}]{dariano_phys.rev.a_45_3646_1992}%
  \BibitemOpen
  \bibfield  {author} {\bibinfo {author} {\bibfnamefont {G.~M.}\ \bibnamefont
  {D'Ariano}}, \bibinfo {author} {\bibfnamefont {L.~R.}\ \bibnamefont
  {Evangelista}},\ and\ \bibinfo {author} {\bibfnamefont {M.}~\bibnamefont
  {Saraceno}},\ }\bibfield  {title} {\bibinfo {title} {Classical and quantum
  structures in the kicked-top model},\ }\href
  {https://doi.org/10.1103/PhysRevA.45.3646} {\bibfield  {journal} {\bibinfo
  {journal} {Phys. Rev. A}\ }\textbf {\bibinfo {volume} {45}},\ \bibinfo
  {pages} {3646} (\bibinfo {year} {1992})}\BibitemShut {NoStop}%
\bibitem [{\citenamefont {Tomsovic}\ and\ \citenamefont
  {Heller}(1993)}]{tomsovic_phys.rev.lett_70_1405_1993}%
  \BibitemOpen
  \bibfield  {author} {\bibinfo {author} {\bibfnamefont {S.}~\bibnamefont
  {Tomsovic}}\ and\ \bibinfo {author} {\bibfnamefont {E.~J.}\ \bibnamefont
  {Heller}},\ }\bibfield  {title} {\bibinfo {title} {Semiclassical construction
  of chaotic eigenstates},\ }\href
  {https://doi.org/10.1103/PhysRevLett.70.1405} {\bibfield  {journal} {\bibinfo
   {journal} {Phys. Rev. Lett.}\ }\textbf {\bibinfo {volume} {70}},\ \bibinfo
  {pages} {1405} (\bibinfo {year} {1993})}\BibitemShut {NoStop}%
\bibitem [{\citenamefont {Revuelta}\ \emph {et~al.}(2020)\citenamefont
  {Revuelta}, \citenamefont {Vergini}, \citenamefont {Benito},\ and\
  \citenamefont {Borondo}}]{revuelta_phys.rev.e_102_042210_2020}%
  \BibitemOpen
  \bibfield  {author} {\bibinfo {author} {\bibfnamefont {F.}~\bibnamefont
  {Revuelta}}, \bibinfo {author} {\bibfnamefont {E.}~\bibnamefont {Vergini}},
  \bibinfo {author} {\bibfnamefont {R.~M.}\ \bibnamefont {Benito}},\ and\
  \bibinfo {author} {\bibfnamefont {F.}~\bibnamefont {Borondo}},\ }\bibfield
  {title} {\bibinfo {title} {Short-periodic-orbit method for excited chaotic
  eigenfunctions},\ }\href {https://doi.org/10.1103/PhysRevE.102.042210}
  {\bibfield  {journal} {\bibinfo  {journal} {Phys. Rev. E}\ }\textbf {\bibinfo
  {volume} {102}},\ \bibinfo {pages} {042210} (\bibinfo {year}
  {2020})}\BibitemShut {NoStop}%
\bibitem [{\citenamefont {Agam}\ and\ \citenamefont
  {Fishman}(1993)}]{agam_j.phys.a.math_26_2113_1993}%
  \BibitemOpen
  \bibfield  {author} {\bibinfo {author} {\bibfnamefont {O.}~\bibnamefont
  {Agam}}\ and\ \bibinfo {author} {\bibfnamefont {S.}~\bibnamefont {Fishman}},\
  }\bibfield  {title} {\bibinfo {title} {Quantum eigenfunctions in terms of
  periodic orbits of chaotic systems},\ }\href
  {https://doi.org/10.1088/0305-4470/26/9/010} {\bibfield  {journal} {\bibinfo
  {journal} {J. Phys. A Math.}\ }\textbf {\bibinfo {volume} {26}},\ \bibinfo
  {pages} {2113} (\bibinfo {year} {1993})}\BibitemShut {NoStop}%
\bibitem [{\citenamefont {Bohigas}\ \emph {et~al.}(1993)\citenamefont
  {Bohigas}, \citenamefont {Tomsovic},\ and\ \citenamefont
  {Ullmo}}]{bohigas_phys.rep_223_43_1993}%
  \BibitemOpen
  \bibfield  {author} {\bibinfo {author} {\bibfnamefont {O.}~\bibnamefont
  {Bohigas}}, \bibinfo {author} {\bibfnamefont {S.}~\bibnamefont {Tomsovic}},\
  and\ \bibinfo {author} {\bibfnamefont {D.}~\bibnamefont {Ullmo}},\ }\bibfield
   {title} {\bibinfo {title} {Manifestations of classical phase space
  structures in quantum mechanics},\ }\href
  {https://doi.org/https://doi.org/10.1016/0370-1573(93)90109-Q} {\bibfield
  {journal} {\bibinfo  {journal} {Phys. rep.}\ }\textbf {\bibinfo {volume}
  {223}},\ \bibinfo {pages} {43} (\bibinfo {year} {1993})}\BibitemShut
  {NoStop}%
\bibitem [{\citenamefont {Wisniacki}\ \emph {et~al.}(2006)\citenamefont
  {Wisniacki}, \citenamefont {Vergini}, \citenamefont {Benito},\ and\
  \citenamefont {Borondo}}]{wisniacki_phys.rev.lett_97_094101_2006}%
  \BibitemOpen
  \bibfield  {author} {\bibinfo {author} {\bibfnamefont {D.~A.}\ \bibnamefont
  {Wisniacki}}, \bibinfo {author} {\bibfnamefont {E.}~\bibnamefont {Vergini}},
  \bibinfo {author} {\bibfnamefont {R.~M.}\ \bibnamefont {Benito}},\ and\
  \bibinfo {author} {\bibfnamefont {F.}~\bibnamefont {Borondo}},\ }\bibfield
  {title} {\bibinfo {title} {Scarring by homoclinic and heteroclinic orbits},\
  }\href {https://doi.org/10.1103/PhysRevLett.97.094101} {\bibfield  {journal}
  {\bibinfo  {journal} {Phys. Rev. Lett.}\ }\textbf {\bibinfo {volume} {97}},\
  \bibinfo {pages} {094101} (\bibinfo {year} {2006})}\BibitemShut {NoStop}%
\bibitem [{\citenamefont {Fromhold}\ \emph {et~al.}(1995)\citenamefont
  {Fromhold}, \citenamefont {Wilkinson}, \citenamefont {Sheard}, \citenamefont
  {Eaves}, \citenamefont {Miao},\ and\ \citenamefont
  {Edwards}}]{Fromhol_phys.rev.lett_75_1142_1995}%
  \BibitemOpen
  \bibfield  {author} {\bibinfo {author} {\bibfnamefont {T.~M.}\ \bibnamefont
  {Fromhold}}, \bibinfo {author} {\bibfnamefont {P.~B.}\ \bibnamefont
  {Wilkinson}}, \bibinfo {author} {\bibfnamefont {F.~W.}\ \bibnamefont
  {Sheard}}, \bibinfo {author} {\bibfnamefont {L.}~\bibnamefont {Eaves}},
  \bibinfo {author} {\bibfnamefont {J.}~\bibnamefont {Miao}},\ and\ \bibinfo
  {author} {\bibfnamefont {G.}~\bibnamefont {Edwards}},\ }\bibfield  {title}
  {\bibinfo {title} {Manifestations of classical chaos in the energy level
  spectrum of a quantum well},\ }\href
  {https://doi.org/10.1103/PhysRevLett.75.1142} {\bibfield  {journal} {\bibinfo
   {journal} {Phys. Rev. Lett.}\ }\textbf {\bibinfo {volume} {75}},\ \bibinfo
  {pages} {1142} (\bibinfo {year} {1995})}\BibitemShut {NoStop}%
\bibitem [{\citenamefont {Wilkinson}\ \emph {et~al.}(1996)\citenamefont
  {Wilkinson}, \citenamefont {Fromhold}, \citenamefont {Eaves}, \citenamefont
  {Sheard}, \citenamefont {Miura},\ and\ \citenamefont
  {Takamasu}}]{Wilkinson_nature_380_608_1996}%
  \BibitemOpen
  \bibfield  {author} {\bibinfo {author} {\bibfnamefont {P.~B.}\ \bibnamefont
  {Wilkinson}}, \bibinfo {author} {\bibfnamefont {T.~M.}\ \bibnamefont
  {Fromhold}}, \bibinfo {author} {\bibfnamefont {L.}~\bibnamefont {Eaves}},
  \bibinfo {author} {\bibfnamefont {F.~W.}\ \bibnamefont {Sheard}}, \bibinfo
  {author} {\bibfnamefont {N.}~\bibnamefont {Miura}},\ and\ \bibinfo {author}
  {\bibfnamefont {T.}~\bibnamefont {Takamasu}},\ }\bibfield  {title} {\bibinfo
  {title} {Observation of 'scarred' wavefunctions in a quantum well with
  chaotic electron dynamics},\ }\href {https://doi.org/10.1038/380608a0}
  {\bibfield  {journal} {\bibinfo  {journal} {Nature (London)}\ }\textbf
  {\bibinfo {volume} {380}},\ \bibinfo {pages} {608} (\bibinfo {year}
  {1996})}\BibitemShut {NoStop}%
\bibitem [{\citenamefont {Narimanov}\ and\ \citenamefont
  {Stone}(1998)}]{narimanov_phys.rev.lett_80_49_1998}%
  \BibitemOpen
  \bibfield  {author} {\bibinfo {author} {\bibfnamefont {E.~E.}\ \bibnamefont
  {Narimanov}}\ and\ \bibinfo {author} {\bibfnamefont {A.~D.}\ \bibnamefont
  {Stone}},\ }\bibfield  {title} {\bibinfo {title} {Origin of strong scarring
  of wave functions in quantum wells in a tilted magnetic field},\ }\href
  {https://doi.org/10.1103/PhysRevLett.80.49} {\bibfield  {journal} {\bibinfo
  {journal} {Phys. Rev. Lett.}\ }\textbf {\bibinfo {volume} {80}},\ \bibinfo
  {pages} {49} (\bibinfo {year} {1998})}\BibitemShut {NoStop}%
\bibitem [{\citenamefont {H{\"o}nig}\ and\ \citenamefont
  {Wintgen}(1989)}]{Honig_phys.rev.a_39_5642_1989}%
  \BibitemOpen
  \bibfield  {author} {\bibinfo {author} {\bibfnamefont {A.}~\bibnamefont
  {H{\"o}nig}}\ and\ \bibinfo {author} {\bibfnamefont {D.}~\bibnamefont
  {Wintgen}},\ }\bibfield  {title} {\bibinfo {title} {Spectral properties of
  strongly perturbed coulomb systems: Fluctuation properties},\ }\href
  {https://doi.org/10.1103/PhysRevA.39.5642} {\bibfield  {journal} {\bibinfo
  {journal} {Phys. Rev. A}\ }\textbf {\bibinfo {volume} {39}},\ \bibinfo
  {pages} {5642} (\bibinfo {year} {1989})}\BibitemShut {NoStop}%
\bibitem [{\citenamefont {Bogomolny}\ \emph {et~al.}(2006)\citenamefont
  {Bogomolny}, \citenamefont {Dietz}, \citenamefont {Friedrich}, \citenamefont
  {Miski-Oglu}, \citenamefont {Richter}, \citenamefont {Sch{\"a}fer},\ and\
  \citenamefont {Schmit}}]{bogomolny_phys.rev.lett_97_254102_2006}%
  \BibitemOpen
  \bibfield  {author} {\bibinfo {author} {\bibfnamefont {E.}~\bibnamefont
  {Bogomolny}}, \bibinfo {author} {\bibfnamefont {B.}~\bibnamefont {Dietz}},
  \bibinfo {author} {\bibfnamefont {T.}~\bibnamefont {Friedrich}}, \bibinfo
  {author} {\bibfnamefont {M.}~\bibnamefont {Miski-Oglu}}, \bibinfo {author}
  {\bibfnamefont {A.}~\bibnamefont {Richter}}, \bibinfo {author} {\bibfnamefont
  {F.}~\bibnamefont {Sch{\"a}fer}},\ and\ \bibinfo {author} {\bibfnamefont
  {C.}~\bibnamefont {Schmit}},\ }\bibfield  {title} {\bibinfo {title} {First
  experimental observation of superscars in a pseudointegrable barrier
  billiard},\ }\href {https://doi.org/10.1103/PhysRevLett.97.254102} {\bibfield
   {journal} {\bibinfo  {journal} {Phys. Rev. Lett.}\ }\textbf {\bibinfo
  {volume} {97}},\ \bibinfo {pages} {254102} (\bibinfo {year}
  {2006})}\BibitemShut {NoStop}%
\bibitem [{\citenamefont {Kim}\ \emph {et~al.}(2002)\citenamefont {Kim},
  \citenamefont {Barth}, \citenamefont {St\"ockmann},\ and\ \citenamefont
  {Bird}}]{kim_phys.rev.b_65_165317_2002}%
  \BibitemOpen
  \bibfield  {author} {\bibinfo {author} {\bibfnamefont {Y.-H.}\ \bibnamefont
  {Kim}}, \bibinfo {author} {\bibfnamefont {M.}~\bibnamefont {Barth}}, \bibinfo
  {author} {\bibfnamefont {H.-J.}\ \bibnamefont {St\"ockmann}},\ and\ \bibinfo
  {author} {\bibfnamefont {J.~P.}\ \bibnamefont {Bird}},\ }\bibfield  {title}
  {\bibinfo {title} {Wave function scarring in open quantum dots: A
  microwave-billiard analog study},\ }\href
  {https://doi.org/10.1103/PhysRevB.65.165317} {\bibfield  {journal} {\bibinfo
  {journal} {Phys. Rev. B}\ }\textbf {\bibinfo {volume} {65}},\ \bibinfo
  {pages} {165317} (\bibinfo {year} {2002})}\BibitemShut {NoStop}%
\bibitem [{\citenamefont {St{\"o}ckmann}\ and\ \citenamefont
  {Stein}(1990)}]{stockman_phys.rev.lett.64.2215_1990}%
  \BibitemOpen
  \bibfield  {author} {\bibinfo {author} {\bibfnamefont {H.-J.}\ \bibnamefont
  {St{\"o}ckmann}}\ and\ \bibinfo {author} {\bibfnamefont {J.}~\bibnamefont
  {Stein}},\ }\bibfield  {title} {\bibinfo {title} {``quantum'' chaos in
  billiards studied by microwave absorption},\ }\href
  {https://doi.org/10.1103/PhysRevLett.64.2215} {\bibfield  {journal} {\bibinfo
   {journal} {Phys. Rev. Lett.}\ }\textbf {\bibinfo {volume} {64}},\ \bibinfo
  {pages} {2215} (\bibinfo {year} {1990})}\BibitemShut {NoStop}%
\bibitem [{\citenamefont {D{\"o}rr}\ \emph {et~al.}(1998)\citenamefont
  {D{\"o}rr}, \citenamefont {St{\"o}ckmann}, \citenamefont {Barth},\ and\
  \citenamefont {Kuhl}}]{dorr_phys.rev.lett_80_1030_1998}%
  \BibitemOpen
  \bibfield  {author} {\bibinfo {author} {\bibfnamefont {U.}~\bibnamefont
  {D{\"o}rr}}, \bibinfo {author} {\bibfnamefont {H.-J.}\ \bibnamefont
  {St{\"o}ckmann}}, \bibinfo {author} {\bibfnamefont {M.}~\bibnamefont
  {Barth}},\ and\ \bibinfo {author} {\bibfnamefont {U.}~\bibnamefont {Kuhl}},\
  }\bibfield  {title} {\bibinfo {title} {Scarred and chaotic field
  distributions in a three-dimensional {S}inai-microwave resonator},\ }\href
  {https://doi.org/10.1103/PhysRevLett.80.1030} {\bibfield  {journal} {\bibinfo
   {journal} {Phys. Rev. Lett.}\ }\textbf {\bibinfo {volume} {80}},\ \bibinfo
  {pages} {1030} (\bibinfo {year} {1998})}\BibitemShut {NoStop}%
\bibitem [{\citenamefont {Sridhar}(1991)}]{Sridhaar_phys.rev.lett_67_785_1991}%
  \BibitemOpen
  \bibfield  {author} {\bibinfo {author} {\bibfnamefont {S.}~\bibnamefont
  {Sridhar}},\ }\bibfield  {title} {\bibinfo {title} {Experimental observation
  of scarred eigenfunctions of chaotic microwave cavities},\ }\href
  {https://doi.org/10.1103/PhysRevLett.67.785} {\bibfield  {journal} {\bibinfo
  {journal} {Phys. Rev. Lett.}\ }\textbf {\bibinfo {volume} {67}},\ \bibinfo
  {pages} {785} (\bibinfo {year} {1991})}\BibitemShut {NoStop}%
\bibitem [{\citenamefont {Stein}\ and\ \citenamefont
  {St{\"o}ckmann}(1992)}]{Stein_phys.rev.lett_68_2867_1992}%
  \BibitemOpen
  \bibfield  {author} {\bibinfo {author} {\bibfnamefont {J.}~\bibnamefont
  {Stein}}\ and\ \bibinfo {author} {\bibfnamefont {H.-J.}\ \bibnamefont
  {St{\"o}ckmann}},\ }\bibfield  {title} {\bibinfo {title} {Experimental
  determination of billiard wave functions},\ }\href
  {https://doi.org/10.1103/PhysRevLett.68.2867} {\bibfield  {journal} {\bibinfo
   {journal} {Phys. Rev. Lett.}\ }\textbf {\bibinfo {volume} {68}},\ \bibinfo
  {pages} {2867} (\bibinfo {year} {1992})}\BibitemShut {NoStop}%
\bibitem [{\citenamefont {N{\"o}ckel}\ and\ \citenamefont
  {Stone}(1997)}]{nockel_nature_385_45_1997}%
  \BibitemOpen
  \bibfield  {author} {\bibinfo {author} {\bibfnamefont {J.}~\bibnamefont
  {N{\"o}ckel}}\ and\ \bibinfo {author} {\bibfnamefont {A.}~\bibnamefont
  {Stone}},\ }\bibfield  {title} {\bibinfo {title} {Ray and wave chaos in
  asymmetric resonant optical cavities},\ }\href
  {https://doi.org/10.1038/385045a0} {\bibfield  {journal} {\bibinfo  {journal}
  {Nature}\ }\textbf {\bibinfo {volume} {385}},\ \bibinfo {pages} {45}
  (\bibinfo {year} {1997})}\BibitemShut {NoStop}%
\bibitem [{\citenamefont {Lee}\ \emph {et~al.}(2002)\citenamefont {Lee},
  \citenamefont {Lee}, \citenamefont {Chang}, \citenamefont {Moon},
  \citenamefont {Kim},\ and\ \citenamefont
  {An}}]{Lee_phys.rev.lett_88_033903_2002}%
  \BibitemOpen
  \bibfield  {author} {\bibinfo {author} {\bibfnamefont {S.-B.}\ \bibnamefont
  {Lee}}, \bibinfo {author} {\bibfnamefont {J.-H.}\ \bibnamefont {Lee}},
  \bibinfo {author} {\bibfnamefont {J.-S.}\ \bibnamefont {Chang}}, \bibinfo
  {author} {\bibfnamefont {H.-J.}\ \bibnamefont {Moon}}, \bibinfo {author}
  {\bibfnamefont {S.~W.}\ \bibnamefont {Kim}},\ and\ \bibinfo {author}
  {\bibfnamefont {K.}~\bibnamefont {An}},\ }\bibfield  {title} {\bibinfo
  {title} {Observation of scarred modes in asymmetrically deformed
  microcylinder lasers},\ }\href
  {https://doi.org/10.1103/PhysRevLett.88.033903} {\bibfield  {journal}
  {\bibinfo  {journal} {Phys. Rev. Lett.}\ }\textbf {\bibinfo {volume} {88}},\
  \bibinfo {pages} {033903} (\bibinfo {year} {2002})}\BibitemShut {NoStop}%
\bibitem [{\citenamefont {Harayama}\ \emph {et~al.}(2003)\citenamefont
  {Harayama}, \citenamefont {Fukushima}, \citenamefont {Davis}, \citenamefont
  {Vaccaro}, \citenamefont {Miyasaka}, \citenamefont {Nishimura},\ and\
  \citenamefont {Aida}}]{Harayama_phys.rev.e_67_015207_2003}%
  \BibitemOpen
  \bibfield  {author} {\bibinfo {author} {\bibfnamefont {T.}~\bibnamefont
  {Harayama}}, \bibinfo {author} {\bibfnamefont {T.}~\bibnamefont {Fukushima}},
  \bibinfo {author} {\bibfnamefont {P.}~\bibnamefont {Davis}}, \bibinfo
  {author} {\bibfnamefont {P.~O.}\ \bibnamefont {Vaccaro}}, \bibinfo {author}
  {\bibfnamefont {T.}~\bibnamefont {Miyasaka}}, \bibinfo {author}
  {\bibfnamefont {T.}~\bibnamefont {Nishimura}},\ and\ \bibinfo {author}
  {\bibfnamefont {T.}~\bibnamefont {Aida}},\ }\bibfield  {title} {\bibinfo
  {title} {Lasing on scar modes in fully chaotic microcavities},\ }\href
  {https://doi.org/10.1103/PhysRevE.67.015207} {\bibfield  {journal} {\bibinfo
  {journal} {Phys. Rev. E}\ }\textbf {\bibinfo {volume} {67}},\ \bibinfo
  {pages} {015207(R)} (\bibinfo {year} {2003})}\BibitemShut {NoStop}%
\bibitem [{\citenamefont {Chinnery}\ and\ \citenamefont
  {Humphrey}(1996)}]{chinnery_phys.rev.e_53_272_1996}%
  \BibitemOpen
  \bibfield  {author} {\bibinfo {author} {\bibfnamefont {P.~A.}\ \bibnamefont
  {Chinnery}}\ and\ \bibinfo {author} {\bibfnamefont {V.~F.}\ \bibnamefont
  {Humphrey}},\ }\bibfield  {title} {\bibinfo {title} {Experimental
  visualization of acoustic resonances within a stadium-shaped cavity},\
  }\href@noop {} {\bibfield  {journal} {\bibinfo  {journal} {Phys. Rev. E}\
  }\textbf {\bibinfo {volume} {53}},\ \bibinfo {pages} {272} (\bibinfo {year}
  {1996})}\BibitemShut {NoStop}%
\bibitem [{\citenamefont {Huang}\ \emph {et~al.}(2009)\citenamefont {Huang},
  \citenamefont {Lai}, \citenamefont {Ferry}, \citenamefont {Goodnick},\ and\
  \citenamefont {Akis}}]{Huang_phys.rev.lett_103_054101_2009}%
  \BibitemOpen
  \bibfield  {author} {\bibinfo {author} {\bibfnamefont {L.}~\bibnamefont
  {Huang}}, \bibinfo {author} {\bibfnamefont {Y.-C.}\ \bibnamefont {Lai}},
  \bibinfo {author} {\bibfnamefont {D.~K.}\ \bibnamefont {Ferry}}, \bibinfo
  {author} {\bibfnamefont {S.~M.}\ \bibnamefont {Goodnick}},\ and\ \bibinfo
  {author} {\bibfnamefont {R.}~\bibnamefont {Akis}},\ }\bibfield  {title}
  {\bibinfo {title} {Relativistic quantum scars},\ }\href
  {https://doi.org/10.1103/PhysRevLett.103.054101} {\bibfield  {journal}
  {\bibinfo  {journal} {Phys. Rev. Lett.}\ }\textbf {\bibinfo {volume} {103}},\
  \bibinfo {pages} {054101} (\bibinfo {year} {2009})}\BibitemShut {NoStop}%
\bibitem [{\citenamefont {Xu}\ \emph {et~al.}(2013)\citenamefont {Xu},
  \citenamefont {Huang}, \citenamefont {Lai},\ and\ \citenamefont
  {Grebogi}}]{Xu_phys.rev.lett_110_064102_2013}%
  \BibitemOpen
  \bibfield  {author} {\bibinfo {author} {\bibfnamefont {H.}~\bibnamefont
  {Xu}}, \bibinfo {author} {\bibfnamefont {L.}~\bibnamefont {Huang}}, \bibinfo
  {author} {\bibfnamefont {Y.-C.}\ \bibnamefont {Lai}},\ and\ \bibinfo {author}
  {\bibfnamefont {C.}~\bibnamefont {Grebogi}},\ }\bibfield  {title} {\bibinfo
  {title} {Chiral scars in chaotic dirac fermion systems},\ }\href
  {https://doi.org/10.1103/PhysRevLett.110.064102} {\bibfield  {journal}
  {\bibinfo  {journal} {Phys. Rev. Lett.}\ }\textbf {\bibinfo {volume} {110}},\
  \bibinfo {pages} {064102} (\bibinfo {year} {2013})}\BibitemShut {NoStop}%
\bibitem [{\citenamefont {Ni}\ \emph {et~al.}(2012)\citenamefont {Ni},
  \citenamefont {Huang}, \citenamefont {Lai},\ and\ \citenamefont
  {Grebogi}}]{xuan_phys.rev.E_86_016702_2012}%
  \BibitemOpen
  \bibfield  {author} {\bibinfo {author} {\bibfnamefont {X.}~\bibnamefont
  {Ni}}, \bibinfo {author} {\bibfnamefont {L.}~\bibnamefont {Huang}}, \bibinfo
  {author} {\bibfnamefont {Y.-C.}\ \bibnamefont {Lai}},\ and\ \bibinfo {author}
  {\bibfnamefont {C.}~\bibnamefont {Grebogi}},\ }\bibfield  {title} {\bibinfo
  {title} {Scarring of dirac fermions in chaotic billiards},\ }\href
  {https://doi.org/10.1103/PhysRevE.86.016702} {\bibfield  {journal} {\bibinfo
  {journal} {Phys. Rev. E}\ }\textbf {\bibinfo {volume} {86}},\ \bibinfo
  {pages} {016702} (\bibinfo {year} {2012})}\BibitemShut {NoStop}%
\bibitem [{\citenamefont {Song}\ \emph {et~al.}(2019)\citenamefont {Song},
  \citenamefont {Li}, \citenamefont {Xu}, \citenamefont {Huang},\ and\
  \citenamefont {Lai}}]{song_phys_rev.research_1_033008_2019}%
  \BibitemOpen
  \bibfield  {author} {\bibinfo {author} {\bibfnamefont {M.-Y.}\ \bibnamefont
  {Song}}, \bibinfo {author} {\bibfnamefont {Z.-Y.}\ \bibnamefont {Li}},
  \bibinfo {author} {\bibfnamefont {H.-Y.}\ \bibnamefont {Xu}}, \bibinfo
  {author} {\bibfnamefont {L.}~\bibnamefont {Huang}},\ and\ \bibinfo {author}
  {\bibfnamefont {Y.-C.}\ \bibnamefont {Lai}},\ }\bibfield  {title} {\bibinfo
  {title} {Quantization of massive dirac billiards and unification of
  nonrelativistic and relativistic chiral quantum scars},\ }\href
  {https://doi.org/10.1103/PhysRevResearch.1.033008} {\bibfield  {journal}
  {\bibinfo  {journal} {Phys. Rev. Research}\ }\textbf {\bibinfo {volume}
  {1}},\ \bibinfo {pages} {033008} (\bibinfo {year} {2019})}\BibitemShut
  {NoStop}%
\bibitem [{\citenamefont {Bernien}\ \emph {et~al.}(2017)\citenamefont
  {Bernien}, \citenamefont {Schwartz}, \citenamefont {Keesling}, \citenamefont
  {Levine}, \citenamefont {Omran}, \citenamefont {Pichler}, \citenamefont
  {Choi}, \citenamefont {Zibrov}, \citenamefont {Endres}, \citenamefont
  {Greiner} \emph {et~al.}}]{bernien_nature_551_579_2017}%
  \BibitemOpen
  \bibfield  {author} {\bibinfo {author} {\bibfnamefont {H.}~\bibnamefont
  {Bernien}}, \bibinfo {author} {\bibfnamefont {S.}~\bibnamefont {Schwartz}},
  \bibinfo {author} {\bibfnamefont {A.}~\bibnamefont {Keesling}}, \bibinfo
  {author} {\bibfnamefont {H.}~\bibnamefont {Levine}}, \bibinfo {author}
  {\bibfnamefont {A.}~\bibnamefont {Omran}}, \bibinfo {author} {\bibfnamefont
  {H.}~\bibnamefont {Pichler}}, \bibinfo {author} {\bibfnamefont
  {S.}~\bibnamefont {Choi}}, \bibinfo {author} {\bibfnamefont {A.~S.}\
  \bibnamefont {Zibrov}}, \bibinfo {author} {\bibfnamefont {M.}~\bibnamefont
  {Endres}}, \bibinfo {author} {\bibfnamefont {M.}~\bibnamefont {Greiner}},
  \emph {et~al.},\ }\bibfield  {title} {\bibinfo {title} {Probing many-body
  dynamics on a 51-atom quantum simulator},\ }\href@noop {} {\bibfield
  {journal} {\bibinfo  {journal} {Nature}\ }\textbf {\bibinfo {volume} {551}},\
  \bibinfo {pages} {579} (\bibinfo {year} {2017})}\BibitemShut {NoStop}%
\bibitem [{\citenamefont {Scherg}\ \emph {et~al.}(2021)\citenamefont {Scherg},
  \citenamefont {Kohlert}, \citenamefont {Sala}, \citenamefont {Pollmann},
  \citenamefont {Hebbe~Madhusudhana}, \citenamefont {Bloch},\ and\
  \citenamefont {Aidelsburger}}]{scherg_nat.commun_12_1_2021}%
  \BibitemOpen
  \bibfield  {author} {\bibinfo {author} {\bibfnamefont {S.}~\bibnamefont
  {Scherg}}, \bibinfo {author} {\bibfnamefont {T.}~\bibnamefont {Kohlert}},
  \bibinfo {author} {\bibfnamefont {P.}~\bibnamefont {Sala}}, \bibinfo {author}
  {\bibfnamefont {F.}~\bibnamefont {Pollmann}}, \bibinfo {author}
  {\bibfnamefont {B.}~\bibnamefont {Hebbe~Madhusudhana}}, \bibinfo {author}
  {\bibfnamefont {I.}~\bibnamefont {Bloch}},\ and\ \bibinfo {author}
  {\bibfnamefont {M.}~\bibnamefont {Aidelsburger}},\ }\bibfield  {title}
  {\bibinfo {title} {Observing non-ergodicity due to kinetic constraints in
  tilted fermi-hubbard chains},\ }\href@noop {} {\bibfield  {journal} {\bibinfo
   {journal} {Nat. Commun.}\ }\textbf {\bibinfo {volume} {12}},\ \bibinfo
  {pages} {1} (\bibinfo {year} {2021})}\BibitemShut {NoStop}%
\bibitem [{\citenamefont {Zhao}\ \emph {et~al.}(2020)\citenamefont {Zhao},
  \citenamefont {Vovrosh}, \citenamefont {Mintert},\ and\ \citenamefont
  {Knolle}}]{zhao_phys.rev.lett_124_160604_2020}%
  \BibitemOpen
  \bibfield  {author} {\bibinfo {author} {\bibfnamefont {H.}~\bibnamefont
  {Zhao}}, \bibinfo {author} {\bibfnamefont {J.}~\bibnamefont {Vovrosh}},
  \bibinfo {author} {\bibfnamefont {F.}~\bibnamefont {Mintert}},\ and\ \bibinfo
  {author} {\bibfnamefont {J.}~\bibnamefont {Knolle}},\ }\bibfield  {title}
  {\bibinfo {title} {Quantum many-body scars in optical lattices},\ }\href
  {https://doi.org/10.1103/PhysRevLett.124.160604} {\bibfield  {journal}
  {\bibinfo  {journal} {Phys. Rev. Lett.}\ }\textbf {\bibinfo {volume} {124}},\
  \bibinfo {pages} {160604} (\bibinfo {year} {2020})}\BibitemShut {NoStop}%
\bibitem [{\citenamefont {Hummel}\ \emph {et~al.}(2023)\citenamefont {Hummel},
  \citenamefont {Richter},\ and\ \citenamefont
  {Schlagheck}}]{Hummel_phys.rev.Lett_130_250402_2023}%
  \BibitemOpen
  \bibfield  {author} {\bibinfo {author} {\bibfnamefont {Q.}~\bibnamefont
  {Hummel}}, \bibinfo {author} {\bibfnamefont {K.}~\bibnamefont {Richter}},\
  and\ \bibinfo {author} {\bibfnamefont {P.}~\bibnamefont {Schlagheck}},\
  }\bibfield  {title} {\bibinfo {title} {Genuine many-body quantum scars along
  unstable modes in bose-hubbard systems},\ }\href
  {https://doi.org/10.1103/PhysRevLett.130.250402} {\bibfield  {journal}
  {\bibinfo  {journal} {Phys. Rev. Lett.}\ }\textbf {\bibinfo {volume} {130}},\
  \bibinfo {pages} {250402} (\bibinfo {year} {2023})}\BibitemShut {NoStop}%
\bibitem [{\citenamefont {Evrard}\ \emph {et~al.}(2024)\citenamefont {Evrard},
  \citenamefont {Pizzi}, \citenamefont {Mistakidis},\ and\ \citenamefont
  {Dag}}]{Evrard_phys.rev.lett_132_020401_2024}%
  \BibitemOpen
  \bibfield  {author} {\bibinfo {author} {\bibfnamefont {B.}~\bibnamefont
  {Evrard}}, \bibinfo {author} {\bibfnamefont {A.}~\bibnamefont {Pizzi}},
  \bibinfo {author} {\bibfnamefont {S.~I.}\ \bibnamefont {Mistakidis}},\ and\
  \bibinfo {author} {\bibfnamefont {C.~B.}\ \bibnamefont {Dag}},\ }\bibfield
  {title} {\bibinfo {title} {Quantum scars and regular eigenstates in a chaotic
  spinor condensate},\ }\href {https://doi.org/10.1103/PhysRevLett.132.020401}
  {\bibfield  {journal} {\bibinfo  {journal} {Phys. Rev. Lett.}\ }\textbf
  {\bibinfo {volume} {132}},\ \bibinfo {pages} {020401} (\bibinfo {year}
  {2024})}\BibitemShut {NoStop}%
\bibitem [{\citenamefont {Turner}\ \emph {et~al.}(2018)\citenamefont {Turner},
  \citenamefont {Michailidis}, \citenamefont {Abanin}, \citenamefont {Serbyn},\
  and\ \citenamefont {Papi{\'c}}}]{turner_nat.phys_14_745_2018}%
  \BibitemOpen
  \bibfield  {author} {\bibinfo {author} {\bibfnamefont {C.~J.}\ \bibnamefont
  {Turner}}, \bibinfo {author} {\bibfnamefont {A.~A.}\ \bibnamefont
  {Michailidis}}, \bibinfo {author} {\bibfnamefont {D.~A.}\ \bibnamefont
  {Abanin}}, \bibinfo {author} {\bibfnamefont {M.}~\bibnamefont {Serbyn}},\
  and\ \bibinfo {author} {\bibfnamefont {Z.}~\bibnamefont {Papi{\'c}}},\
  }\bibfield  {title} {\bibinfo {title} {Weak ergodicity breaking from quantum
  many-body scars},\ }\href@noop {} {\bibfield  {journal} {\bibinfo  {journal}
  {Nat. Phys.}\ }\textbf {\bibinfo {volume} {14}},\ \bibinfo {pages} {745}
  (\bibinfo {year} {2018})}\BibitemShut {NoStop}%
\bibitem [{\citenamefont {Ho}\ \emph {et~al.}(2019)\citenamefont {Ho},
  \citenamefont {Choi}, \citenamefont {Pichler},\ and\ \citenamefont
  {Lukin}}]{Ho_phys.rev.lett_122_040603_2019}%
  \BibitemOpen
  \bibfield  {author} {\bibinfo {author} {\bibfnamefont {W.~W.}\ \bibnamefont
  {Ho}}, \bibinfo {author} {\bibfnamefont {S.}~\bibnamefont {Choi}}, \bibinfo
  {author} {\bibfnamefont {H.}~\bibnamefont {Pichler}},\ and\ \bibinfo {author}
  {\bibfnamefont {M.~D.}\ \bibnamefont {Lukin}},\ }\bibfield  {title} {\bibinfo
  {title} {Periodic orbits, entanglement, and quantum many-body scars in
  constrained models: Matrix product state approach},\ }\href
  {https://doi.org/10.1103/PhysRevLett.122.040603} {\bibfield  {journal}
  {\bibinfo  {journal} {Phys. Rev. Lett.}\ }\textbf {\bibinfo {volume} {122}},\
  \bibinfo {pages} {040603} (\bibinfo {year} {2019})}\BibitemShut {NoStop}%
\bibitem [{\citenamefont {Serbyn}\ \emph {et~al.}(2021)\citenamefont {Serbyn},
  \citenamefont {Abanin},\ and\ \citenamefont
  {Papi{\'c}}}]{serbyn_nat.phys_17_675_2021}%
  \BibitemOpen
  \bibfield  {author} {\bibinfo {author} {\bibfnamefont {M.}~\bibnamefont
  {Serbyn}}, \bibinfo {author} {\bibfnamefont {D.~A.}\ \bibnamefont {Abanin}},\
  and\ \bibinfo {author} {\bibfnamefont {Z.}~\bibnamefont {Papi{\'c}}},\
  }\bibfield  {title} {\bibinfo {title} {Quantum many-body scars and weak
  breaking of ergodicity},\ }\href@noop {} {\bibfield  {journal} {\bibinfo
  {journal} {Nat. Phys.}\ }\textbf {\bibinfo {volume} {17}},\ \bibinfo {pages}
  {675} (\bibinfo {year} {2021})}\BibitemShut {NoStop}%
\bibitem [{\citenamefont {Luukko}\ \emph {et~al.}(2016)\citenamefont {Luukko},
  \citenamefont {Drury}, \citenamefont {Klales}, \citenamefont {Kaplan},
  \citenamefont {Heller},\ and\ \citenamefont
  {R{\"a}s{\"a}nen}}]{Luukko_sci.rep_6_37656_2016}%
  \BibitemOpen
  \bibfield  {author} {\bibinfo {author} {\bibfnamefont {P.~J.~J.}\
  \bibnamefont {Luukko}}, \bibinfo {author} {\bibfnamefont {B.}~\bibnamefont
  {Drury}}, \bibinfo {author} {\bibfnamefont {A.}~\bibnamefont {Klales}},
  \bibinfo {author} {\bibfnamefont {L.}~\bibnamefont {Kaplan}}, \bibinfo
  {author} {\bibfnamefont {E.~J.}\ \bibnamefont {Heller}},\ and\ \bibinfo
  {author} {\bibfnamefont {E.}~\bibnamefont {R{\"a}s{\"a}nen}},\ }\bibfield
  {title} {\bibinfo {title} {Strong quantum scarring by local impurities},\
  }\href {https://doi.org/10.1038/srep37656} {\bibfield  {journal} {\bibinfo
  {journal} {Sci. Rep.}\ }\textbf {\bibinfo {volume} {6}},\ \bibinfo {pages}
  {37656} (\bibinfo {year} {2016})}\BibitemShut {NoStop}%
\bibitem [{\citenamefont {Keski-Rahkonen}\ \emph {et~al.}(2017)\citenamefont
  {Keski-Rahkonen}, \citenamefont {Luukko}, \citenamefont {Kaplan},
  \citenamefont {Heller},\ and\ \citenamefont
  {R{\"a}s{\"a}nen}}]{keski-rahkonen_phys.rev.b_97_094204_2017}%
  \BibitemOpen
  \bibfield  {author} {\bibinfo {author} {\bibfnamefont {J.}~\bibnamefont
  {Keski-Rahkonen}}, \bibinfo {author} {\bibfnamefont {P.~J.~J.}\ \bibnamefont
  {Luukko}}, \bibinfo {author} {\bibfnamefont {L.}~\bibnamefont {Kaplan}},
  \bibinfo {author} {\bibfnamefont {E.~J.}\ \bibnamefont {Heller}},\ and\
  \bibinfo {author} {\bibfnamefont {E.}~\bibnamefont {R{\"a}s{\"a}nen}},\
  }\bibfield  {title} {\bibinfo {title} {Controllable quantum scars in
  semiconductor quantum dots},\ }\href
  {https://doi.org/10.1103/PhysRevB.96.094204} {\bibfield  {journal} {\bibinfo
  {journal} {Phys. Rev. B}\ }\textbf {\bibinfo {volume} {96}},\ \bibinfo
  {pages} {094204} (\bibinfo {year} {2017})}\BibitemShut {NoStop}%
\bibitem [{\citenamefont {Keski-Rahkonen}\ \emph
  {et~al.}(2019{\natexlab{a}})\citenamefont {Keski-Rahkonen}, \citenamefont
  {Luukko}, \citenamefont {{\AA}berg},\ and\ \citenamefont
  {R{\"a}s{\"a}nen}}]{keski-rahkonen_j.phys.conden.matter_31_105301_2019}%
  \BibitemOpen
  \bibfield  {author} {\bibinfo {author} {\bibfnamefont {J.}~\bibnamefont
  {Keski-Rahkonen}}, \bibinfo {author} {\bibfnamefont {P.~J.~J.}\ \bibnamefont
  {Luukko}}, \bibinfo {author} {\bibfnamefont {S.}~\bibnamefont {{\AA}berg}},\
  and\ \bibinfo {author} {\bibfnamefont {E.}~\bibnamefont {R{\"a}s{\"a}nen}},\
  }\bibfield  {title} {\bibinfo {title} {Effects of scarring on quantum chaos
  in disordered quantum wells},\ }\href
  {https://doi.org/10.1088/1361-648x/aaf9fb} {\bibfield  {journal} {\bibinfo
  {journal} {J. Phys.: Condens. Matter}\ }\textbf {\bibinfo {volume} {31}},\
  \bibinfo {pages} {105301} (\bibinfo {year} {2019}{\natexlab{a}})}\BibitemShut
  {NoStop}%
\bibitem [{\citenamefont {Keski-Rahkonen}\ \emph
  {et~al.}(2019{\natexlab{b}})\citenamefont {Keski-Rahkonen}, \citenamefont
  {Ruhanen}, \citenamefont {Heller},\ and\ \citenamefont
  {R{\"a}s{\"a}nen}}]{keski-rahkonen_phys.rev.lett_123_214101_2019}%
  \BibitemOpen
  \bibfield  {author} {\bibinfo {author} {\bibfnamefont {J.}~\bibnamefont
  {Keski-Rahkonen}}, \bibinfo {author} {\bibfnamefont {A.}~\bibnamefont
  {Ruhanen}}, \bibinfo {author} {\bibfnamefont {E.~J.}\ \bibnamefont
  {Heller}},\ and\ \bibinfo {author} {\bibfnamefont {E.}~\bibnamefont
  {R{\"a}s{\"a}nen}},\ }\bibfield  {title} {\bibinfo {title} {Quantum lissajous
  scars},\ }\href {https://doi.org/10.1103/PhysRevLett.123.214101} {\bibfield
  {journal} {\bibinfo  {journal} {Phys. Rev. Lett.}\ }\textbf {\bibinfo
  {volume} {123}},\ \bibinfo {pages} {214101} (\bibinfo {year}
  {2019}{\natexlab{b}})}\BibitemShut {NoStop}%
\bibitem [{\citenamefont {Luukko}\ and\ \citenamefont
  {Rost}(2017)}]{Luukko_phys.rev.lett_119_203001_2017}%
  \BibitemOpen
  \bibfield  {author} {\bibinfo {author} {\bibfnamefont {P.~J.~J.}\
  \bibnamefont {Luukko}}\ and\ \bibinfo {author} {\bibfnamefont {J.-M.}\
  \bibnamefont {Rost}},\ }\bibfield  {title} {\bibinfo {title} {Polyatomic
  trilobite rydberg molecules in a dense random gas},\ }\href
  {https://doi.org/10.1103/PhysRevLett.119.203001} {\bibfield  {journal}
  {\bibinfo  {journal} {Phys. Rev. Lett.}\ }\textbf {\bibinfo {volume} {119}},\
  \bibinfo {pages} {203001} (\bibinfo {year} {2017})}\BibitemShut {NoStop}%
\bibitem [{\citenamefont {Keski-Rahkonen}\ \emph {et~al.}(2024)\citenamefont
  {Keski-Rahkonen}, \citenamefont {Graf},\ and\ \citenamefont
  {Heller}}]{antiscarring}%
  \BibitemOpen
  \bibfield  {author} {\bibinfo {author} {\bibfnamefont {J.}~\bibnamefont
  {Keski-Rahkonen}}, \bibinfo {author} {\bibfnamefont {A.}~\bibnamefont
  {Graf}},\ and\ \bibinfo {author} {\bibfnamefont {E.}~\bibnamefont {Heller}},\
  }\bibfield  {title} {\bibinfo {title} {Antiscarring in chaotic quantum
  wells},\ }\href@noop {} {\bibfield  {journal} {\bibinfo  {journal} {arXiv
  preprint arXiv:2403.18081}\ } (\bibinfo {year} {2024})}\BibitemShut {NoStop}%
\bibitem [{\citenamefont {O'Connor}\ and\ \citenamefont
  {Heller}(1988)}]{OConnor_phys.rev.lett_61_2288_1988}%
  \BibitemOpen
  \bibfield  {author} {\bibinfo {author} {\bibfnamefont {P.~W.}\ \bibnamefont
  {O'Connor}}\ and\ \bibinfo {author} {\bibfnamefont {E.~J.}\ \bibnamefont
  {Heller}},\ }\bibfield  {title} {\bibinfo {title} {Quantum localization for a
  strongly classically chaotic system},\ }\href
  {https://doi.org/10.1103/PhysRevLett.61.2288} {\bibfield  {journal} {\bibinfo
   {journal} {Phys. Rev. Lett.}\ }\textbf {\bibinfo {volume} {61}},\ \bibinfo
  {pages} {2288} (\bibinfo {year} {1988})}\BibitemShut {NoStop}%
\bibitem [{\citenamefont {Halonen}\ \emph {et~al.}(1996)\citenamefont
  {Halonen}, \citenamefont {Hyv\"onen}, \citenamefont {Pietil\"ainen},\ and\
  \citenamefont {Chakraborty}}]{Halonen_Phys.Rev.B_53_6971_1996}%
  \BibitemOpen
  \bibfield  {author} {\bibinfo {author} {\bibfnamefont {V.}~\bibnamefont
  {Halonen}}, \bibinfo {author} {\bibfnamefont {P.}~\bibnamefont {Hyv\"onen}},
  \bibinfo {author} {\bibfnamefont {P.}~\bibnamefont {Pietil\"ainen}},\ and\
  \bibinfo {author} {\bibfnamefont {T.}~\bibnamefont {Chakraborty}},\
  }\bibfield  {title} {\bibinfo {title} {Effects of scattering centers on the
  energy spectrum of a quantum dot},\ }\href
  {https://doi.org/10.1103/PhysRevB.53.6971} {\bibfield  {journal} {\bibinfo
  {journal} {Phys. Rev. B}\ }\textbf {\bibinfo {volume} {53}},\ \bibinfo
  {pages} {6971} (\bibinfo {year} {1996})}\BibitemShut {NoStop}%
\bibitem [{\citenamefont {R\"as\"anen}\ \emph {et~al.}(2004)\citenamefont
  {R\"as\"anen}, \citenamefont {K\"onemann}, \citenamefont {Haug},
  \citenamefont {Puska},\ and\ \citenamefont
  {Nieminen}}]{Rasanen_Phys.Rev.B_70_115308_2004}%
  \BibitemOpen
  \bibfield  {author} {\bibinfo {author} {\bibfnamefont {E.}~\bibnamefont
  {R\"as\"anen}}, \bibinfo {author} {\bibfnamefont {J.}~\bibnamefont
  {K\"onemann}}, \bibinfo {author} {\bibfnamefont {R.~J.}\ \bibnamefont
  {Haug}}, \bibinfo {author} {\bibfnamefont {M.~J.}\ \bibnamefont {Puska}},\
  and\ \bibinfo {author} {\bibfnamefont {R.~M.}\ \bibnamefont {Nieminen}},\
  }\bibfield  {title} {\bibinfo {title} {Impurity effects in quantum dots:
  Toward quantitative modeling},\ }\href
  {https://doi.org/10.1103/PhysRevB.70.115308} {\bibfield  {journal} {\bibinfo
  {journal} {Phys. Rev. B}\ }\textbf {\bibinfo {volume} {70}},\ \bibinfo
  {pages} {115308} (\bibinfo {year} {2004})}\BibitemShut {NoStop}%
\bibitem [{\citenamefont {G\"u\ifmmode~\mbox{\c{c}}\else \c{c}\fi{}l\"u}\ \emph
  {et~al.}(2003)\citenamefont {G\"u\ifmmode~\mbox{\c{c}}\else \c{c}\fi{}l\"u},
  \citenamefont {Wang},\ and\ \citenamefont
  {Guo}}]{Guclu_Phys.Rev.B_68_035304_2003}%
  \BibitemOpen
  \bibfield  {author} {\bibinfo {author} {\bibfnamefont {A.~D.}\ \bibnamefont
  {G\"u\ifmmode~\mbox{\c{c}}\else \c{c}\fi{}l\"u}}, \bibinfo {author}
  {\bibfnamefont {J.-S.}\ \bibnamefont {Wang}},\ and\ \bibinfo {author}
  {\bibfnamefont {H.}~\bibnamefont {Guo}},\ }\bibfield  {title} {\bibinfo
  {title} {Disordered quantum dots: A diffusion quantum monte carlo study},\
  }\href {https://doi.org/10.1103/PhysRevB.68.035304} {\bibfield  {journal}
  {\bibinfo  {journal} {Phys. Rev. B}\ }\textbf {\bibinfo {volume} {68}},\
  \bibinfo {pages} {035304} (\bibinfo {year} {2003})}\BibitemShut {NoStop}%
\bibitem [{\citenamefont {Mendoza}\ and\ \citenamefont
  {Schulz}(2003)}]{mendoza_phys.rev.b_68_205302_2003}%
  \BibitemOpen
  \bibfield  {author} {\bibinfo {author} {\bibfnamefont {M.}~\bibnamefont
  {Mendoza}}\ and\ \bibinfo {author} {\bibfnamefont {P.~A.}\ \bibnamefont
  {Schulz}},\ }\bibfield  {title} {\bibinfo {title} {Wave-function mapping
  conditions in open quantum dot structures},\ }\href@noop {} {\bibfield
  {journal} {\bibinfo  {journal} {Phys. Rev. B}\ }\textbf {\bibinfo {volume}
  {68}},\ \bibinfo {pages} {205302} (\bibinfo {year} {2003})}\BibitemShut
  {NoStop}%
\bibitem [{\citenamefont {Zozoulenko}\ \emph {et~al.}(1998)\citenamefont
  {Zozoulenko}, \citenamefont {Sachrajda}, \citenamefont {Zawadzki},
  \citenamefont {Berggren}, \citenamefont {Feng},\ and\ \citenamefont
  {Wasilewski}}]{zozoulenko_phys.rev.b_5810597_1998}%
  \BibitemOpen
  \bibfield  {author} {\bibinfo {author} {\bibfnamefont {I.~V.}\ \bibnamefont
  {Zozoulenko}}, \bibinfo {author} {\bibfnamefont {A.~S.}\ \bibnamefont
  {Sachrajda}}, \bibinfo {author} {\bibfnamefont {P.}~\bibnamefont {Zawadzki}},
  \bibinfo {author} {\bibfnamefont {K.-F.}\ \bibnamefont {Berggren}}, \bibinfo
  {author} {\bibfnamefont {Y.}~\bibnamefont {Feng}},\ and\ \bibinfo {author}
  {\bibfnamefont {Z.}~\bibnamefont {Wasilewski}},\ }\bibfield  {title}
  {\bibinfo {title} {Conductance fluctuations in a rectangular dot at constant
  magnetic fields},\ }\href@noop {} {\bibfield  {journal} {\bibinfo  {journal}
  {Phys. Rev. B}\ }\textbf {\bibinfo {volume} {58}},\ \bibinfo {pages} {10597}
  (\bibinfo {year} {1998})}\BibitemShut {NoStop}%
\bibitem [{\citenamefont {Jullien}\ \emph {et~al.}(2014)\citenamefont
  {Jullien}, \citenamefont {Roulleau}, \citenamefont {Roche}, \citenamefont
  {Cavanna}, \citenamefont {Jin},\ and\ \citenamefont
  {Glattli}}]{jullien2_Nature_514_603_2014}%
  \BibitemOpen
  \bibfield  {author} {\bibinfo {author} {\bibfnamefont {T.}~\bibnamefont
  {Jullien}}, \bibinfo {author} {\bibfnamefont {P.}~\bibnamefont {Roulleau}},
  \bibinfo {author} {\bibfnamefont {B.}~\bibnamefont {Roche}}, \bibinfo
  {author} {\bibfnamefont {A.}~\bibnamefont {Cavanna}}, \bibinfo {author}
  {\bibfnamefont {Y.}~\bibnamefont {Jin}},\ and\ \bibinfo {author}
  {\bibfnamefont {D.}~\bibnamefont {Glattli}},\ }\bibfield  {title} {\bibinfo
  {title} {Quantum tomography of an electron},\ }\href@noop {} {\bibfield
  {journal} {\bibinfo  {journal} {Nature}\ }\textbf {\bibinfo {volume} {514}},\
  \bibinfo {pages} {603} (\bibinfo {year} {2014})}\BibitemShut {NoStop}%
\bibitem [{\citenamefont {Ge}\ \emph {et~al.}(2024)\citenamefont {Ge},
  \citenamefont {Graf}, \citenamefont {Keski-Rahkonen}, \citenamefont
  {Slizovskiy}, \citenamefont {Polizogopoulos}, \citenamefont {Taniguchi},
  \citenamefont {Watanabe}, \citenamefont {Van~Haren}, \citenamefont
  {Lederman}, \citenamefont {Fal'ko} \emph {et~al.}}]{ge2024direct}%
  \BibitemOpen
  \bibfield  {author} {\bibinfo {author} {\bibfnamefont {Z.}~\bibnamefont
  {Ge}}, \bibinfo {author} {\bibfnamefont {A.~M.}\ \bibnamefont {Graf}},
  \bibinfo {author} {\bibfnamefont {J.}~\bibnamefont {Keski-Rahkonen}},
  \bibinfo {author} {\bibfnamefont {S.}~\bibnamefont {Slizovskiy}}, \bibinfo
  {author} {\bibfnamefont {P.}~\bibnamefont {Polizogopoulos}}, \bibinfo
  {author} {\bibfnamefont {T.}~\bibnamefont {Taniguchi}}, \bibinfo {author}
  {\bibfnamefont {K.}~\bibnamefont {Watanabe}}, \bibinfo {author}
  {\bibfnamefont {R.}~\bibnamefont {Van~Haren}}, \bibinfo {author}
  {\bibfnamefont {D.}~\bibnamefont {Lederman}}, \bibinfo {author}
  {\bibfnamefont {V.~I.}\ \bibnamefont {Fal'ko}}, \emph {et~al.},\ }\bibfield
  {title} {\bibinfo {title} {Direct visualization of relativistic quantum scars
  in graphene quantum dots},\ }\href
  {https://doi.org/10.1038/s41586-024-08190-6} {\bibfield  {journal} {\bibinfo
  {journal} {Nature}\ }\textbf {\bibinfo {volume} {635}},\ \bibinfo {pages}
  {841} (\bibinfo {year} {2024})}\BibitemShut {NoStop}%
\bibitem [{\citenamefont {Hirose}\ and\ \citenamefont
  {Wingreen}(2002)}]{hirose_phys.rev.B_65_193305_2002}%
  \BibitemOpen
  \bibfield  {author} {\bibinfo {author} {\bibfnamefont {K.}~\bibnamefont
  {Hirose}}\ and\ \bibinfo {author} {\bibfnamefont {N.~S.}\ \bibnamefont
  {Wingreen}},\ }\bibfield  {title} {\bibinfo {title} {Ground-state energy and
  spin in disordered quantum dots},\ }\href@noop {} {\bibfield  {journal}
  {\bibinfo  {journal} {Phys. Rev. B}\ }\textbf {\bibinfo {volume} {65}},\
  \bibinfo {pages} {193305} (\bibinfo {year} {2002})}\BibitemShut {NoStop}%
\bibitem [{\citenamefont {Hirose}\ \emph {et~al.}(2001)\citenamefont {Hirose},
  \citenamefont {Zhou},\ and\ \citenamefont
  {Wingreen}}]{hirose_phys.Rev.B_63_075301_2001}%
  \BibitemOpen
  \bibfield  {author} {\bibinfo {author} {\bibfnamefont {K.}~\bibnamefont
  {Hirose}}, \bibinfo {author} {\bibfnamefont {F.}~\bibnamefont {Zhou}},\ and\
  \bibinfo {author} {\bibfnamefont {N.~S.}\ \bibnamefont {Wingreen}},\
  }\bibfield  {title} {\bibinfo {title} {Density-functional theory of
  spin-polarized disordered quantum dots},\ }\href@noop {} {\bibfield
  {journal} {\bibinfo  {journal} {Phys. Rev. B}\ }\textbf {\bibinfo {volume}
  {63}},\ \bibinfo {pages} {075301} (\bibinfo {year} {2001})}\BibitemShut
  {NoStop}%
\bibitem [{\citenamefont {Luukko}\ and\ \citenamefont
  {Räsänen}(2013)}]{LuukkoP.J.J.2013Itpc}%
  \BibitemOpen
  \bibfield  {author} {\bibinfo {author} {\bibfnamefont {P.}~\bibnamefont
  {Luukko}}\ and\ \bibinfo {author} {\bibfnamefont {E.}~\bibnamefont
  {Räsänen}},\ }\bibfield  {title} {\bibinfo {title} {Imaginary time
  propagation code for large-scale two-dimensional eigenvalue problems in
  magnetic fields},\ }\href@noop {} {\bibfield  {journal} {\bibinfo  {journal}
  {Computer physics communications}\ }\textbf {\bibinfo {volume} {184}},\
  \bibinfo {pages} {769} (\bibinfo {year} {2013})}\BibitemShut {NoStop}%
\end{thebibliography}%

\end{document}